\documentclass[aps,a4paper,showkeys,nofootinbib,longbibliography,twocolumn]{revtex4-2}
\usepackage{braket}
\usepackage[utf8]{inputenc}
\usepackage{filecontents}
\usepackage{natbib}
\usepackage{amsmath,amssymb,bm,amsthm}
\usepackage{subfigure}
\usepackage{graphicx}
\usepackage{dcolumn}
\usepackage{bm}
\usepackage{booktabs}
\usepackage{color}
\newcounter{one}
\usepackage{url}

\usepackage{textcase}

\usepackage{colortbl}
\usepackage{tabularx}
\usepackage{verbatim}
\usepackage{multirow}

\usepackage[T1]{fontenc} 
\usepackage{lmodern}
\usepackage{bbm}
\usepackage[utf8]{inputenc}
\usepackage{amsfonts}
\usepackage{array}

\usepackage{bbm}
\usepackage{enumitem}
\usepackage{umoline}
\usepackage[usenames,svgnames]{xcolor}
\usepackage{natbib}
\usepackage[hyperindex,breaklinks]{hyperref}
\hypersetup{
	colorlinks=true, 
	linkcolor=Navy, 
	citecolor=Navy, 
	filecolor=Navy,
	urlcolor=Navy,  
	runcolor=cyan,
}
\usepackage{graphicx}
\usepackage{amsfonts}
\usepackage{amssymb}
\usepackage{amsmath}
\usepackage{bbm}
\usepackage{enumitem}
\usepackage{xcolor}
\usepackage{dsfont}
\usepackage{mathrsfs}

\newcommand{\half}[1]{{ \rm h}}

\usepackage{yhmath}

\def\beq{\begin{equation}}
	\def\eeq{\end{equation}}
\def\nbeq{\begin{equation*}}
	\def\neeq{\end{equation*}}
\def\<{\langle}
\def\>{\rangle}

\def\Tr{{\rm Tr}}

\def\Tr{{\rm Tr}}

\newtheorem{theorem}{Theorem}
\newtheorem{subtheorem}{Subtheorem}
\newtheorem{lemma}{Lemma}
\newtheorem{corollary}{Corollary}

\newtheorem{proposition}[subtheorem]{Proposition}

\definecolor{bulgarianrose}{rgb}{0.28, 0.02, 0.03}
\definecolor{rosewood}{rgb}{0.4, 0.0, 0.04}

\newcommand{\br}[1]{\left( #1 \right)}
\newcommand{\brr}[1]{\left[ #1 \right]}
\newcommand{\brrr}[1]{\left\{ #1 \right\}}
\newcommand{\norm}[1]{\left \|  #1 \right \|}

\newcommand{\abs}[1]{\left| #1 \right|}
  
\usepackage{yhmath}

\def\Tr{{\rm Tr}}

\def\Tr{{\rm Tr}}

\newcommand{\ii}{\mathrm{i}}
\newcommand{\ee}{\mathrm{e}}
\newcommand{\dd}{\mathrm{d}}
\newcommand{\cO}{\mathcal{O}}
\newcommand{\id}{\mathds{1}}
\newcommand{\sq}{\mathrm{sq}} 
\DeclareMathOperator{\Ran}{Ran}

\begin{document}

\title{Tunable Ground-State Entanglement in Fully Connected Quantum Systems: \\ From Logarithmic to Volume-Law Scaling}

\author{Donghoon Kim}
\email{donghoon.kim@riken.jp}
\affiliation{Analytical Quantum Complexity RIKEN Hakubi Research Team, RIKEN Center for Quantum Computing (RQC), Wako, Saitama 351-0198, Japan}

\author{Tomotaka Kuwahara}
\email{tomotaka.kuwahara@riken.jp}
\affiliation{Analytical Quantum Complexity RIKEN Hakubi Research Team, RIKEN Center for Quantum Computing (RQC), Wako, Saitama 351-0198, Japan}
\affiliation{RIKEN Pioneering Research Institute (PRI), Wako, Saitama 351-0198, Japan}

\begin{abstract}
	Entanglement scaling is a central probe of the complexity of quantum many-body ground states, yet once geometric locality is absent, it can depend sensitively on the structure and normalization of interactions.
	For fully connected Hamiltonians with the natural Kac normalization, nondegenerate gapped ground states are known to obey a logarithmic entanglement bound.
	Here we ask how the ground-state entanglement changes when this interaction normalization is continuously relaxed away from the Kac scale.
	We answer this question by constructing an explicit all-to-all bipartite Hamiltonian of $n$ qubits with Fourier pairing amplitudes and pairwise interaction strength proportional to $n^{-\kappa}$.
	For sufficiently small fixed coupling, we prove that every pure ground state has entanglement entropy  $\Theta(n^{2-2\kappa} \log n)$ for $1/2<\kappa\leq1$ and $\Theta(n)$ at $\kappa=1/2$.
	We further show that the ground state is unique and separated from the rest of the spectrum by a system-size-independent gap for $3/4 \leq \kappa \leq 1$, while weak-coupling analysis and numerical calculations suggest that the gap remains open also for $1/2 \leq \kappa < 3/4$.
	Thus, varying only the interaction normalization yields a controlled hierarchy from logarithmic scaling at the natural Kac normalization $\kappa=1$ to increasingly strong subextensive entanglement and ultimately to a volume law, showing that even an arbitrarily weak relaxation of this normalization precludes a universal logarithmic entanglement bound.
\end{abstract}

\maketitle

\section{Introduction}
\label{sec:introduction}

The scaling of ground-state entanglement in quantum many-body systems is fundamentally constrained by the structure and strength of their interactions~\cite{vidal2003entanglement,calabrese2004entanglement,amico2008entanglement,RevModPhys.81.865,laflorencie2016quantum}.
In geometrically local systems, these constraints are closely tied to entanglement area laws~\cite{RevModPhys.82.277,cramer2006entanglement,Hastings_2007,masanes2009area,PhysRevB.85.195145,arad2013area,brandao2013area,brandao2015exponential,AnshuAradGosset2022}, which bound ground-state entanglement entropy by the size of the subsystem boundary and play a central role in understanding gapped phases and the effectiveness of tensor-network methods~\cite{white1992density,white1993density,verstraete2008mps,schollwock2011density,orus2014practical,landau2015polynomial,arad2017rigorousRG,cirac2021matrix}.
Even when geometric locality is relaxed to allow long-range interactions, entanglement area laws can persist when the interactions decay sufficiently rapidly with distance~\cite{gong2017entanglement,Kuwahara2020arealaw,LiuYiZhouZou2025,KimKuwahara2026}.
Without such spatial restrictions, however, it becomes less clear what general constraints on ground-state entanglement remain, and explicit counterexamples on general interaction graphs show that analogous area-law behavior can fail~\cite{aharonov2014local}.

Among systems with nonlocal interactions, fully connected and related all-to-all Hamiltonians form an important class, in which each degree of freedom can interact with an extensive number of others.
Such systems arise in a wide range of contexts, including collective spin models~\cite{Lipkin1965}, spin glasses~\cite{Sherrington1975}, SYK-type models~\cite{Sachdev1993}, and several quantum-simulation platforms~\cite{Blatt2012,Monroe2013}.
These systems exhibit a wide range of ground-state entanglement scalings.
For example, the Lipkin--Meshkov--Glick model exhibits at most logarithmic bipartite entanglement~\cite{PhysRevA.71.064101}, whereas SYK-type models and transverse-field quantum Sherrington--Kirkpatrick models can exhibit much stronger ground-state entanglement, including volume-law scaling~\cite{PhysRevB.94.035135,PhysRevLett.129.220401}.
These contrasting behaviors show that connectivity alone does not determine the scaling of ground-state entanglement.

Recent work~\cite{KimKuwahara2024} has established a universal logarithmic upper bound on ground-state entanglement in fully connected systems of $n$ qudits, despite the absence of geometric locality.
More precisely, when the on-site terms are bounded and the pairwise interactions scale as $\mathcal{O}(1/n)$, every nondegenerate ground state with a uniform spectral gap obeys an $\mathcal{O}(\log n)$ entanglement bound across arbitrary bipartitions.
This interaction scaling corresponds to the natural Kac normalization~\cite{KacUhlenbeckHemmer1963,defenu2023long}, under which the total interaction strength incident on each site remains bounded as the system size increases.

More recently, Ref.~\cite{KimKuwahara2026Collectivity} extended ground-state entanglement bounds to a broader class of geometrically nonlocal systems with Kac-normalized power-law interactions decaying as $r^{-\alpha}$ with distance $r$ on $D$-dimensional lattices, proving an $\mathcal{O}(\log n)$ bound across arbitrary bipartitions for $\alpha<D/2$.
Beyond these upper bounds, the work also provided explicit constructions of uniformly gapped long-range Hamiltonians exhibiting matching $\Omega(\log n)$ ground-state entanglement in this regime, thereby establishing the optimality of logarithmic scaling, including in the fully connected case $\alpha=0$.
These bounds and lower-bound constructions are obtained within the Kac-normalized regime, leaving open whether the normalization itself is essential for the resulting entanglement constraints or merely a sufficient condition.
For fully connected systems, this leads to the question of whether the $1/n$ interaction scaling marks a sharp threshold for universal logarithmic entanglement bounds, even in the presence of a uniform spectral gap.

In this work, we investigate how interaction normalization governs the scaling of ground-state entanglement in fully connected quantum systems. 
We construct an explicit family of $n$-qubit Hamiltonians with Fourier-structured all-to-all bipartite couplings, whose pairwise strengths scale as $n^{-\kappa}$ for $1/2 \leq \kappa \leq 1$, with $\kappa = 1$ corresponding to the standard Kac normalization.
For sufficiently small fixed coupling, we establish explicit two-sided finite-size bounds on the entanglement entropy of every normalized pure ground state.
These bounds imply $S_{A}=\Theta(n^{2-2\kappa}\log n)$ for $1/2<\kappa\leq1$ and $S_{A}=\Theta(n)$ at $\kappa=1/2$.
These results establish a hierarchy of entanglement scaling laws within a single family of fully connected Hamiltonians and reveal the fragility of logarithmic scaling at the Kac normalization: even an arbitrarily small fixed reduction of the normalization exponent leads to a polynomial enhancement of entanglement. 
Across this family, the total interaction strength incident on each site grows as $n^{1-\kappa}$ for $\kappa < 1$, while the ground-state energy density remains bounded.

Under the same weak-coupling condition, we further establish that the ground state is unique and separated from the rest of the spectrum by a system-size-independent gap for $3/4 \leq \kappa \leq 1$.
Thus, uniformly gapped ground states can exhibit superlogarithmic entanglement under arbitrarily small fixed departures from the Kac normalization, establishing the sharpness of the $1/n$ interaction scaling for universal logarithmic entanglement bounds.
To further characterize the entanglement entropy and spectral gap, we derive quantitative weak-coupling predictions using quadratic Holstein--Primakoff theory and spin perturbation theory, respectively, and compare them with density-matrix renormalization group (DMRG) calculations.
These complementary analyses are consistent with the rigorously established entanglement scaling and provide evidence that the spectral gap may remain open also for $1/2\leq\kappa<3/4$, beyond the range of the uniform-gap proof.

This manuscript is organized as follows.
Section~\ref{sec:model} introduces the Fourier-paired Hamiltonian and its interaction normalization.
Section~\ref{sec:rigorous} presents the rigorous entropy and spectral-gap results.
Section~\ref{sec:weak-coupling} derives the weak-coupling formulas, and Sec.~\ref{sec:numerics} presents the numerical comparisons.
Section~\ref{sec:discussion} discusses the implications of our results and the remaining open questions.
The appendices contain the detailed proofs and calculations.

\section{Setup and Model}
\label{sec:model}

We consider a system of $n=2N$ qubits on a site set $\Lambda=A\sqcup B$, where $A=\{i_{0},\ldots,i_{N-1}\}$ and $B=\{j_{0},\ldots,j_{N-1}\}$.
The corresponding spin Hilbert spaces are $\mathcal{H}_{A}:=(\mathbb{C}^{2})^{\otimes N}$ and $\mathcal{H}_{B}:=(\mathbb{C}^{2})^{\otimes N}$, with the full Hilbert space $\mathcal{H}:=\mathcal{H}_{A}\otimes\mathcal{H}_{B}$.
At each site, we use the computational basis $\{\ket{0},\ket{1}\}$ and define $n_{x}:=\br{\ket{1}\bra{1}}_{x}$, $\sigma_{x}^{+}:=\br{\ket{1}\bra{0}}_{x}$, and $\sigma_{x}^{-}:=\br{\ket{0}\bra{1}}_{x}$.
The excitation-number operators on the two subsystems are
\begin{equation}
	\mathcal{N}_{A} := \sum_{a=0}^{N-1}n_{i_{a}},\qquad \mathcal{N}_{B} := \sum_{b=0}^{N-1}n_{j_{b}}.
	\label{eq:number-operators}
\end{equation}
The total excitation number is $\mathcal{N}:=\mathcal{N}_{A}+\mathcal{N}_{B}$.
The vacuum is $\ket{\mathbf{0}}:=\ket{0}^{\otimes 2N}$, so that $\mathcal{N}\ket{\mathbf{0}}=0$.

Let $\omega_{N} := \ee^{\frac{2\pi\ii}{N}}$ and let $F_{N}$ denote the normalized discrete Fourier matrix with entries $(F_{N})_{ab} := N^{-\frac{1}{2}} \omega_{N}^{ab}$, so that $F_{N}^{\dagger}F_{N} = F_{N}F_{N}^{\dagger} = \id_{N}$.
The corresponding pair-creation and pair-annihilation operators are
\begin{equation}
	P_{N}^{\dagger} := \sum_{a,b=0}^{N-1} (F_{N})_{ab} \sigma_{i_{a}}^{+} \sigma_{j_{b}}^{+},\qquad P_{N} := (P_{N}^{\dagger})^{\dagger}.
	\label{eq:P-def}
\end{equation}
For a coupling parameter $g\geq0$, we consider the Hamiltonian
\begin{equation}
	H_{N}(g) := \mathcal{N} - g\bigl(P_{N}^{\dagger}+P_{N}\bigr).
	\label{eq:fourier-hamiltonian-g}
\end{equation}
The coefficient of $\mathcal{N}$ is set to one, so that a single-spin excitation has energy one in the noninteracting limit $g=0$.
We denote the ground-state energy of $H_{N}(g)$ by $E_{N}(g)$ and its spectral gap by $\Delta_{N}(g)$.
For $\gamma \geq 0$ and $1/2 \leq \kappa \leq 1$, set $g_{N,\kappa} := \gamma N^{\frac{1}{2}-\kappa}$ and write $H_{N,\kappa} := H_{N}(g_{N,\kappa})$.
Explicitly,
\begin{equation}
	H_{N,\kappa} = \mathcal{N} - \frac{\gamma}{N^{\kappa}} \sum_{a,b=0}^{N-1} \left(\omega_{N}^{ab} \sigma_{i_{a}}^{+} \sigma_{j_{b}}^{+} + \omega_{N}^{-ab} \sigma_{i_{a}}^{-} \sigma_{j_{b}}^{-} \right).
	\label{eq:fourier-hamiltonian}
\end{equation}
Thus, $H_{N,\kappa}$ is an all-to-all bipartite pairing Hamiltonian: every site in $A$ couples to every site in $B$, with pairwise coupling magnitude $\gamma N^{-\kappa}$.
Each interaction creates or annihilates one excitation on each side, rather than exchanging an excitation between the two subsystems.
Consequently, the excitation-number difference $\mathcal{D}:=\mathcal{N}_{A}-\mathcal{N}_{B}$ is conserved.
Moreover, since each pairing term changes the total excitation number by two, the total excitation parity $\Pi:=(-1)^{\mathcal{N}}$ is also conserved.
Therefore, we have
\begin{equation}
	[H_{N}(g),\mathcal{D}] = [H_{N}(g),\Pi] = 0.
	\label{eq:conserved-quantities}
\end{equation}

Let $h_{i_{a}j_{b}}$ denote the two-site interaction on $\{i_{a},j_{b}\}$ in Eq.~\eqref{eq:fourier-hamiltonian}.
It acts only on $\operatorname{span}\{\ket{00},\ket{11}\}$, with eigenvalues $\pm\gamma N^{-\kappa}$ on that subspace, and hence $\norm{h_{i_{a}j_{b}}}=\gamma N^{-\kappa}$.
The total incident interaction strength at a fixed site $i_{a}$ is therefore
\begin{equation}
	\sum_{b=0}^{N-1} \norm{h_{i_{a}j_{b}}} = \gamma N^{1-\kappa}.
	\label{eq:incident-interaction}
\end{equation}
Thus, $\kappa=1$ is the natural Kac normalization: each pair interaction has norm $\gamma/N=2\gamma/n$, while the total incident interaction strength remains of order one~\cite{KacUhlenbeckHemmer1963,defenu2023long}.
For every $\kappa<1$, by contrast, the incident interaction strength grows as $N^{1-\kappa}$.

We focus on the bipartite entanglement across the partition $A|B$.
For a normalized pure state $\ket{\psi}$, let $\rho_{A}:=\Tr_{B}\br{\ket{\psi}\bra{\psi}}$ and define
\begin{equation}
	S_{A}(\ket{\psi}) := -\Tr\br{\rho_{A}\log\rho_{A}}.
	\label{eq:entanglement-entropy}
\end{equation}
Our goal is to determine the system-size scaling of $S_{A}$ as a function of $\kappa$.
All logarithms are natural logarithms.

More generally, the entanglement and spectral-gap results established below remain valid if the Fourier matrix $F_{N}$ is replaced by any $N\times N$ unitary matrix $U_{N}$, since the corresponding proofs use only $U_{N}^{\dagger}U_{N} = U_{N}U_{N}^{\dagger}=\id_{N}$.
For concreteness, however, we focus throughout on the Fourier matrix.
Its flat-entry property $|(F_{N})_{ab}| = N^{-\frac{1}{2}}$ ensures that every pairwise coupling is nonzero and of equal magnitude, so the model retains the fully connected interaction structure described above.

\section{Rigorous Entanglement Scaling and Spectral Gap}
\label{sec:rigorous}

We now present two main results for $H_{N,\kappa}$ in Eq.~\eqref{eq:fourier-hamiltonian}: explicit bounds and asymptotic scaling laws for the ground-state entanglement entropy, and a criterion for ground-state uniqueness and a uniform spectral gap.
We then combine these results to establish uniformly gapped ground states with superlogarithmic entanglement.

To state the entanglement bounds, define
\begin{equation}
	r(g):=\frac{g}{1+\sqrt{1+g^{2}}},
	\label{eq:r-def}
\end{equation}
and write $r_{N,\kappa} := r(g_{N,\kappa})$.
We also set
\begin{equation}
	L_{N,\kappa} := \min\brrr{N, \left\lfloor\frac{1}{2r_{N,\kappa}^{2}}\right\rfloor}.
	\label{eq:L-def}
\end{equation}

Our first main result establishes the asymptotic scaling of the ground-state entanglement entropy, together with explicit two-sided finite-size bounds.
The result applies to every normalized pure ground state without assuming ground-state uniqueness or a spectral gap.

\begin{theorem}
	\label{thm:finite-size-main}
	Assume $0 < \gamma \leq 1/64$, $1/2 \leq \kappa \leq 1$, and $N \geq 4096$.
	For every normalized pure ground state $\ket{\Omega_{N,\kappa}}$ of $H_{N,\kappa}$ in Eq.~\eqref{eq:fourier-hamiltonian},
	\begin{equation}
		\frac{Nr_{N,\kappa}^{2}}{80} \log L_{N,\kappa} \leq S_{A}(\ket{\Omega_{N,\kappa}}) \leq Nh_{2}(4r_{N,\kappa}^{2}),
		\label{eq:finite-size-two-sided}
	\end{equation}
	where $h_{2}(x):= -x\log x - (1-x)\log(1-x)$.
	Consequently, for fixed $\gamma$ and $\kappa$, the entanglement entropy satisfies
	\begin{equation}
		S_{A}(\ket{\Omega_{N,\kappa}}) = 
		\begin{cases}
			\Theta_{\gamma,\kappa} \br{N^{2-2\kappa} \log N}, & 1/2<\kappa\leq1,\\[1mm]
			\Theta_{\gamma}(N), & \kappa=1/2.
		\end{cases}
		\label{eq:asymptotic-entropy}
	\end{equation}
\end{theorem}

The asymptotic scaling in Eq.~\eqref{eq:asymptotic-entropy} follows directly from the finite-size bounds in Eq.~\eqref{eq:finite-size-two-sided}.
For fixed $\kappa>1/2$, $g_{N,\kappa}\to0$ as $N\to\infty$, and
\begin{equation}
	r_{N,\kappa}^{2} = \frac{\gamma^{2}}{4}N^{1-2\kappa}\brr{1+\cO\br{\gamma^{2}N^{1-2\kappa}}}.
	\label{eq:r-asymptotic}
\end{equation}
If $1/2<\kappa<1$, Eq.~\eqref{eq:L-def} gives $L_{N,\kappa} = \lfloor(2r_{N,\kappa}^{2})^{-1}\rfloor$ for sufficiently large $N$, and hence $\log L_{N,\kappa} = (2\kappa-1)\log N+\cO_{\gamma}(1)$.
At $\kappa=1$, $Nr_{N,\kappa}^{2} < \gamma^{2}/4 < 1/2$, so $L_{N,\kappa}=N$.
Together with Eq.~\eqref{eq:r-asymptotic}, the lower bound in Eq.~\eqref{eq:finite-size-two-sided} therefore gives $S_{A}(\ket{\Omega_{N,\kappa}}) = \Omega_{\gamma,\kappa}(N^{2-2\kappa}\log N)$ for $1/2<\kappa\leq1$.

For the upper bound, Eq.~\eqref{eq:r-asymptotic} gives $\log \br{\ee/(4r_{N,\kappa}^{2})} = (2\kappa-1) \log N+\cO_{\gamma}(1)$.
Using $h_{2}(x) \leq x \log\br{\ee/x}$ in Eq.~\eqref{eq:finite-size-two-sided}, we obtain
\begin{align}
	S_{A}(\ket{\Omega_{N,\kappa}}) &\leq 4Nr_{N,\kappa}^{2} \log\br{\frac{\ee}{4r_{N,\kappa}^{2}}} \notag \\
	&= \cO_{\gamma,\kappa}\br{N^{2-2\kappa}\log N}.
\end{align}
The matching lower and upper bounds establish the first line of Eq.~\eqref{eq:asymptotic-entropy}.

At $\kappa=1/2$, $r_{N,\kappa}=r(\gamma)>0$ is independent of $N$, while $L_{N,\kappa}$ eventually becomes a fixed integer greater than one.
Both bounds in Eq.~\eqref{eq:finite-size-two-sided} are therefore proportional to $N$, proving the second line of Eq.~\eqref{eq:asymptotic-entropy}.

Our second main result establishes ground-state uniqueness and a uniform spectral gap under a finite-size condition on the coupling.
Applied to $g=g_{N,\kappa}$, this yields a system-size-independent gap for $3/4\leq\kappa\leq1$.

\begin{theorem}
	\label{thm:uniform-gap}
	Let $N\geq1$ and assume $0 < g \leq1/64$.
	If $Ng^{4} < 3/2$, then $H_{N}(g)$ in Eq.~\eqref{eq:fourier-hamiltonian-g} has a unique ground state and its spectral gap satisfies
	\begin{equation}
		\Delta_{N}(g)\geq\frac{3}{4}-\frac{1}{2}Ng^{4}.
		\label{eq:finite-size-spin-gap}
	\end{equation}
	In particular, if $Ng^{4}\leq1/2$, then $\Delta_{N}(g)\geq1/2$.
	For the Hamiltonian $H_{N,\kappa}=H_{N}(g_{N,\kappa})$ in Eq.~\eqref{eq:fourier-hamiltonian}, assume $0<\gamma\leq1/64$ and $3/4\leq\kappa\leq1$.
	Then $Ng_{N,\kappa}^{4} \leq \gamma^{4} < 1/2$ for every $N\geq1$, and its spectral gap $\Delta_{N,\kappa} := \Delta_{N}(g_{N,\kappa})$ satisfies
	\begin{equation}
		\Delta_{N,\kappa} \geq \frac{3}{4} - \frac{\gamma^{4}}{2}N^{3-4\kappa} \geq \frac{3}{4} - \frac{\gamma^{4}}{2} > \frac{1}{2}.
		\label{eq:kappa-uniform-gap}
	\end{equation}
	The unique ground state lies in $\ker\mathcal{D}$ and has even parity.
\end{theorem}

Combining the two main theorems, we obtain uniformly gapped ground states with superlogarithmic entanglement under arbitrarily small fixed departures from the Kac normalization.

\begin{corollary}
	\label{cor:gapped-entanglement}
	Fix $0 < \gamma \leq 1/64$ independently of $N$.
	For every fixed $3/4 \leq \kappa < 1$, the Hamiltonian $H_{N,\kappa}$ has a unique uniformly gapped ground state whose bipartite entanglement entropy satisfies $S_{A}=\Theta_{\gamma,\kappa}(n^{2-2\kappa}\log n)$.
	At the natural Kac normalization $\kappa=1$, the ground state remains unique and uniformly gapped, with $S_{A} = \Theta_{\gamma}(\log n)$.
	Thus, the Fourier family saturates the logarithmic system-size dependence of the general entanglement upper bound at the Kac normalization, while arbitrarily small fixed reductions of the normalization exponent produce uniformly gapped ground states with superlogarithmic entanglement.
\end{corollary}

The two main theorems apply over different ranges of $\kappa$.
The entanglement scaling is established throughout $1/2\leq\kappa\leq1$, whereas the uniform-gap result covers $3/4\leq\kappa\leq1$.
This difference follows from $Ng_{N,\kappa}^{4}=\gamma^{4}N^{3-4\kappa}$.
For $\kappa>3/4$, this quantity tends to zero as $N\to\infty$, while at $\kappa=3/4$ it remains equal to the $N$-independent constant $\gamma^{4}$.
For $\kappa<3/4$, the present nonperturbative estimate is insufficient to establish an $N$-uniform positive spectral gap.
This limitation motivates the complementary weak-coupling and numerical analyses in Secs.~\ref{sec:weak-coupling} and \ref{sec:numerics}.

Our construction also constrains the spectral-gap dependence of general entanglement bounds at the standard Kac normalization.
The existing general entanglement upper bound for Kac-normalized fully connected systems scales as $\cO((\bar{g}/\Delta)^{3}\log n)$, where $\bar{g}$ bounds the sum of the operator norms of all Hamiltonian terms acting on any single site~\cite{KimKuwahara2024}.
To assess the sharpness of this gap dependence, we rescale $H_{N,\kappa}$ for $3/4\leq\kappa<1$ as $\widetilde{H}_{N,\kappa}:=N^{\kappa-1}H_{N,\kappa}$, so that its pairwise interactions have strength $\gamma/N$.
This rescaling preserves the ground state and its entanglement entropy, while the spectral gap becomes $\widetilde{\Delta}_{N,\kappa}=N^{\kappa-1}\Delta_{N,\kappa}$.
Since $n=2N$ and the rescaled on-site strength is $N^{\kappa-1}\leq1$, we may take $\bar{g}=1+2\gamma$ for $\widetilde{H}_{N,\kappa}$.
Using $\Delta_{N,\kappa}>1/2$ from Theorem~\ref{thm:uniform-gap}, we obtain $(\bar{g}/\widetilde{\Delta}_{N,\kappa})^{2}<4\bar{g}^{2}N^{2-2\kappa}$, which, together with Theorem~\ref{thm:finite-size-main}, implies
\begin{equation}
	S_{A}=\Omega_{\gamma,\kappa}\left((\bar{g}/\widetilde{\Delta}_{N,\kappa})^{2}\log n\right).
	\label{eq:kac-gap-lower}
\end{equation}
Consequently, no universal entanglement upper bound of the form $\cO((\bar{g}/\Delta)^{p}\log n)$ can hold for $p<2$.
Thus, the optimal inverse-gap exponent $p_{\mathrm{opt}}$ in such bounds satisfies $2 \leq p_{\mathrm{opt}} \leq 3$.

The remainder of this section establishes the finite-size entanglement bounds in Theorem~\ref{thm:finite-size-main} and proves the uniform-gap result in Theorem~\ref{thm:uniform-gap}.

\subsection{Exact One-Sided Bosonic Dilation}
\label{subsec:exact-dilation}

To exploit the unitary pairing structure, we enlarge only the $A$ subsystem to a bosonic Fock space while keeping the $B$ subsystem unchanged.
This construction is not a bosonic approximation: the physical spin Hamiltonian is recovered exactly by compression to the embedded hard-core subspace.
The enlargement instead allows the unitary pairing matrix to be absorbed into a rotation of the bosonic modes, yielding a Hamiltonian on the enlarged Hilbert space that decomposes into $N$ commuting boson--spin channels.
Each channel admits an explicit decomposition into invariant sectors, allowing its ground state and spectral gap to be determined exactly and providing a controlled reference problem for the bounds below.

Let $\mathcal{F}_{A}$ be the bosonic Fock space of $N$ modes with creation operators $a_{a}^{\dagger}$.
Let $\mathcal{K}_{A}\subset\mathcal{F}_{A}$ denote the hard-core subspace spanned by occupation-number states with occupation zero or one in each original mode $a_{a}$.
For each subset $R\subseteq\{0,\ldots,N-1\}$, let $\ket{R}_{A}$ denote the computational-basis state in which precisely the spins indexed by $R$ are excited, namely $\ket{R}_{A}:=\bigl(\prod_{a\in R}\sigma_{i_{a}}^{+}\bigr)\ket{0}^{\otimes N}$.
The local isometry $J_{A}:\mathcal{H}_{A}\to\mathcal{K}_{A}$ is defined on these basis states by
\begin{equation}
	J_{A} \ket{R}_{A} := \prod_{a\in R} a_{a}^{\dagger} \ket{0}_{\mathcal{F}_{A}}.
	\label{eq:hardcore-isometry}
\end{equation}
Thus, each spin excitation is identified with a single boson in the corresponding original mode, and the hard-core constraint is precisely the restriction to at most one boson per mode.
The bosonic operators need not preserve $\mathcal{K}_{A}$.
Nevertheless, their compression to this subspace reproduces the spin operators exactly:
$J_{A}^{\dagger}a_{a}^{\dagger}J_{A}=\sigma_{i_{a}}^{+}$, $J_{A}^{\dagger}a_{a}J_{A}=\sigma_{i_{a}}^{-}$, and $J_{A}^{\dagger}a_{a}^{\dagger}a_{a}J_{A}=n_{i_{a}}$.

After embedding the $A$ subsystem into the bosonic Fock space, the unitary pairing matrix can be absorbed into a change of bosonic modes.
Define
\begin{equation}
	d_{b}^{\dagger} := \sum_{a=0}^{N-1}(F_{N})_{ab}a_{a}^{\dagger}, \quad d_{b} := \sum_{a=0}^{N-1} \overline{(F_{N})_{ab}}a_{a}.
	\label{eq:rotated-bosons}
\end{equation}
Unitarity of $F_{N}$ gives $[d_{b},d_{b'}^{\dagger}]=\delta_{bb'}$, so the $d_{b}$ form canonical bosonic modes.
In this rotated basis, each mode $d_{b}$ couples only to the corresponding physical spin $j_{b}$.
We introduce the following reference Hamiltonian on $\mathcal{F}_{A} \otimes \mathcal{H}_{B}$:
\begin{equation}
	H_{N}^{\mathrm{ref}}(g) := \sum_{b=0}^{N-1}h_{b}(g),
	\label{eq:reference-hamiltonian}
\end{equation}
where the individual channel Hamiltonians are
\begin{equation}
	h_{b}(g) := d_{b}^{\dagger} d_{b} + n_{j_{b}} - g\br{d_{b}^{\dagger} \sigma_{j_{b}}^{+} + d_{b}\sigma_{j_{b}}^{-}}.
	\label{eq:reference-channel}
\end{equation}
Thus, $H_{N}^{\mathrm{ref}}(g)$ decomposes into $N$ identical boson--spin problems acting on distinct mode--spin pairs.

The enlarged reference problem is related to the physical spin Hamiltonian $H_{N}(g)$ in Eq.~\eqref{eq:fourier-hamiltonian-g} by an exact compression.
Writing $J:=J_{A}\otimes\id_{B}$ and $\ket{\widetilde{\psi}}:=J\ket{\psi}$, we have
\begin{align}
	H_{N}(g) &= J^{\dagger} H_{N}^{\mathrm{ref}}(g) J,
	\label{eq:exact-compression} \\
	\bra{\psi}H_{N}(g)\ket{\psi} &= \bra{\widetilde{\psi}}H_{N}^{\mathrm{ref}}(g)\ket{\widetilde{\psi}}.
	\label{eq:exact-compression-expectation}
\end{align}
Hence, every spin-state energy expectation is reproduced exactly by the reference Hamiltonian on the embedded state.
Moreover, the embedding and the bosonic mode rotation act only within subsystem $A$ and therefore preserve the nonzero Schmidt coefficients across $A|B$.
The reference Hamiltonian $H_{N}^{\mathrm{ref}}(g)$ in Eq.~\eqref{eq:reference-hamiltonian}, however, acts on the larger Hilbert space $\mathcal{F}_{A}\otimes\mathcal{H}_{B}$, whose states are not restricted to the hard-core subspace $\mathcal{K}_{A}\otimes\mathcal{H}_{B}$.
Its spectrum therefore need not coincide with that of the physical spin Hamiltonian $H_{N}(g)$ in Eq.~\eqref{eq:fourier-hamiltonian-g}.
Instead, its exactly solvable ground state and spectral gap provide the reference quantities used below to bound the energy, spectral gap, and entanglement properties of the physical spin model.

For $0\leq r<1$, define the normalized state associated with the $b$th boson--spin channel by
\begin{equation}
	\ket{\varphi_{r}}_{b} := \frac{\ket{0,0}_{b} + r\ket{1,1}_{b}}{\sqrt{1+r^{2}}}.
	\label{eq:reference-local-ground}
\end{equation}
Here $\ket{x,y}_{b}:=\ket{x}_{d_{b}}\otimes\ket{y}_{j_{b}}$, where $x\in\mathbb{N}_{0}$ is the occupation number of the bosonic mode $d_{b}$ and $y\in\{0,1\}$ denotes the state of the physical spin $j_{b}$.
For $0\leq g\leq1/4$, diagonalizing the channel Hamiltonian $h_{b}(g)$ in Eq.~\eqref{eq:reference-channel} shows that $\ket{\varphi_{r(g)}}_{b}$, with $r(g)$ defined in Eq.~\eqref{eq:r-def}, is its unique normalized ground state, with ground energy $e_{\mathrm{ref}}(g) := 1-\sqrt{1+g^{2}}$.
These quantities satisfy
\begin{equation}
	g = \frac{2r(g)}{1-r(g)^{2}}, \quad e_{\mathrm{ref}}(g) = -\frac{2r(g)^{2}}{1-r(g)^{2}}.
	\label{eq:g-r-relations}
\end{equation}
Comparison with the remaining eigenvalues obtained from the same diagonalization shows that, for $0\leq g\leq1/4$, the local spectral gap above $\ket{\varphi_{r(g)}}_{b}$ is at least $3/4$, as established in Lemma~\ref{lem:reference-spectrum} of Appendix~\ref{app:dilation}, where the full channel spectrum is also derived.
Since the terms in Eq.~\eqref{eq:reference-hamiltonian} act on distinct mode--spin pairs, their ground states tensorize.
Define
\begin{align}
	\ket{\Phi_{r,N}} &:= \bigotimes_{b=0}^{N-1}\ket{\varphi_{r}}_{b},
	\label{eq:reference-product-state}\\
	Q_{b}(r) &:= \id-\ket{\varphi_{r}}\bra{\varphi_{r}}_{b}.
	\label{eq:reference-projectors}
\end{align}
At $r=r(g)$, the local gap estimate gives
$h_{b}(g)\geq e_{\mathrm{ref}}(g)\id+(3/4)Q_{b}(r(g))$ for each channel, and summing over $b$ yields
\begin{align}
	H_{N}^{\mathrm{ref}}(g) &\geq Ne_{\mathrm{ref}}(g)+\frac{3}{4}\sum_{b}Q_{b}(r(g)) \notag \\
	&\geq Ne_{\mathrm{ref}}(g) + \frac{3}{4} \br{\id-\ket{\Phi_{r(g),N}}\bra{\Phi_{r(g),N}}}.
	\label{eq:reference-gap}
\end{align}
Since the projectors $Q_{b}(r(g))$ act on distinct channels, they commute, and in their common eigenbasis one has $\sum_{b} Q_{b}(r(g)) \geq \id - \prod_{b}(\id-Q_{b}(r(g))) = \id - \ket{\Phi_{r(g),N}} \bra{\Phi_{r(g),N}}$.
This proves the second inequality in Eq.~\eqref{eq:reference-gap}.
The first inequality controls the sum of the local excitation probabilities and will be used in the entanglement bounds below, whereas the second gives a lower bound on the excitation energy relative to the reference ground state and will be used in the proof of ground-state uniqueness and the physical spin gap.

\subsection{Variational Control of the Reference-Energy Defect}
\label{subsec:energy-defect}

Because the reference Hamiltonian $H_{N}^{\mathrm{ref}}(g)$ in Eq.~\eqref{eq:reference-hamiltonian} is minimized over a larger Hilbert space, Eq.~\eqref{eq:exact-compression-expectation} gives a lower bound on the physical ground-state energy.
Recall that $E_{N}(g)$ denotes the ground-state energy of $H_{N}(g)$.
We define the resulting reference-energy defect by
\begin{equation}
	\delta_{N}(g) := E_{N}(g) - Ne_{\mathrm{ref}}(g).
	\label{eq:energy-defect-definition}
\end{equation}
Then $\delta_{N}(g)\geq0$.
The following proposition gives the crucial complementary estimate that the reference-energy defect is bounded by $\cO(Ng^{4})$ uniformly in $N$.

\begin{proposition}
	\label{prop:energy-defect}
	For every $N\geq1$ and every $0\leq g\leq1/64$,
	\begin{equation}
		0 \leq \delta_{N}(g) \leq \frac{1}{2}Ng^{4}.
		\label{eq:energy-defect-bound}
	\end{equation}
\end{proposition}

To prove the upper bound in Eq.~\eqref{eq:energy-defect-bound}, we use a variational state constructed entirely in the finite-dimensional spin Hilbert space.
The pair operators $P_{N}^{\dagger}$ and $P_{N}$ defined in Eq.~\eqref{eq:P-def} preserve the excitation-number difference $\mathcal{D}:=\mathcal{N}_{A}-\mathcal{N}_{B}$, as stated in Eq.~\eqref{eq:conserved-quantities}.
Starting from the vacuum, it is therefore natural to work in $\mathcal{H}_{\mathrm{eq}} := \ker\mathcal{D}$.
On $\mathcal{H}_{\mathrm{eq}}$, the excitation numbers on the two subsystems coincide.
We denote their common restriction by
$T:=\mathcal{N}_{A}|_{\mathcal{H}_{\mathrm{eq}}}=\mathcal{N}_{B}|_{\mathcal{H}_{\mathrm{eq}}}$.
Since the vacuum is the ground state at $g=0$ and $P_{N}^{\dagger}$ creates excitations only in pairs, we introduce the one-parameter variational ansatz
\begin{equation}
	\ket{\psi_{s}} := \ee^{s(P_{N}^{\dagger} - P_{N})}\ket{\mathbf{0}}.
	\label{eq:trial-flow-main}
\end{equation}
The resulting states remain in $\mathcal{H}_{\mathrm{eq}}$, where $\mathcal{N}=2T$.
Writing $m(s):=\bra{\psi_{s}}T\ket{\psi_{s}}$, differentiation gives $m'(s)=\bra{\psi_{s}}(P_{N}^{\dagger}+P_{N})\ket{\psi_{s}}$.
Since $H_{N}(g)=\mathcal{N}-g(P_{N}^{\dagger}+P_{N})$, its variational expectation along Eq.~\eqref{eq:trial-flow-main} is $2m(s)-gm'(s)$.
It is therefore sufficient to obtain an upper bound on $m(s)$ and a lower bound on $m'(s)$.

Differentiating once more gives
$m''(s)=2\bra{\psi_{s}}[P_{N},P_{N}^{\dagger}]\ket{\psi_{s}}$.
The desired bounds on $m(s)$ and $m'(s)$ can therefore be obtained from upper and lower bounds on the pair commutator.
Using the unitarity of $F_{N}$ and the spin commutation relations, Appendix~\ref{app:variational} establishes
\begin{equation}
	N\id-6T\leq[P_{N},P_{N}^{\dagger}] \leq N\id \quad \text{on }\mathcal{H}_{\mathrm{eq}}.
	\label{eq:pair-commutator-main}
\end{equation}
Taking expectations in $\ket{\psi_{s}}$ gives
\begin{equation}
	2N-12m(s)\leq m''(s)\leq2N.
	\label{eq:flow-differential-main}
\end{equation}
With $m(0)=m'(0)=0$, integrating the upper bound in Eq.~\eqref{eq:flow-differential-main} twice gives $m(s)\leq Ns^{2}$ for $s\geq0$.
Combining this with the lower bound in Eq.~\eqref{eq:flow-differential-main} and integrating once gives $m'(s) \geq 2Ns-4Ns^{3}$ for $s \geq 0$.
The detailed derivations are given in Appendix~\ref{app:variational}.

Evaluating the variational energy at $s=g/2$ and using the bounds above gives
\begin{align}
	E_{N}(g) &\leq 2m\br{\frac{g}{2}}-gm'\br{\frac{g}{2}}  \leq -\frac{1}{2}Ng^{2}+\frac{1}{2}Ng^{4}.
	\label{eq:variational-energy-main}
\end{align}
Combining Eq.~\eqref{eq:variational-energy-main} with $e_{\mathrm{ref}}(g)=1-\sqrt{1+g^{2}}\geq-g^{2}/2$ yields the upper bound in Eq.~\eqref{eq:energy-defect-bound}, and hence Proposition~\ref{prop:energy-defect}.
Together with $E_{N}(g)\geq Ne_{\mathrm{ref}}(g)\geq-Ng^{2}/2$, the same estimate also gives
\begin{equation}
	-\frac{1}{2}Ng^{2} \leq E_{N}(g) \leq -\frac{1}{2}Ng^{2}+\frac{1}{2}Ng^{4}.
	\label{eq:uniform-spin-energy}
\end{equation}

Setting $g = g_{N,\kappa} = \gamma N^{\frac{1}{2}-\kappa}$ in Eq.~\eqref{eq:uniform-spin-energy} gives
\begin{equation}
	E_{N}(g_{N,\kappa}) = -\frac{\gamma^{2}}{2}N^{2-2\kappa} + \cO\br{\gamma^{4}N^{3-4\kappa}}.
	\label{eq:spin-energy-N-rigorous}
\end{equation}
For fixed $\gamma$ and $\kappa>1/2$, the error is smaller than the leading term by a factor of order $N^{1-2\kappa}$.
In particular, $E_{N}(g_{N,\kappa})/(2N)\to0$ for $\kappa>1/2$, whereas at $\kappa=1/2$ Eq.~\eqref{eq:uniform-spin-energy} gives a negative ground-state energy proportional to $N$ for the stated small couplings.

\subsection{Uniqueness and Gap from the Same Energy Estimate}
\label{subsec:gap-proof}

The reference Hamiltonian $H_{N}^{\mathrm{ref}}(g)$ in Eq.~\eqref{eq:reference-hamiltonian} has the fixed spectral gap in Eq.~\eqref{eq:reference-gap} above its product ground state $\ket{\Phi_{r(g),N}}$, while Proposition~\ref{prop:energy-defect} bounds how far the physical ground energy can lie above the reference ground energy.
The combination of these two facts gives a direct min--max comparison for the physical spin spectrum and proves Theorem~\ref{thm:uniform-gap} without a size-dependent perturbation expansion.

Let $E_{N,0}(g) \leq E_{N,1}(g) \leq \cdots$ denote the eigenvalues of the physical spin Hamiltonian $H_{N}(g)$ in Eq.~\eqref{eq:fourier-hamiltonian-g}, arranged in nondecreasing order and counted with multiplicity, so that $E_{N,0}(g) = E_{N}(g)$.
With $\ket{\Phi_{r(g),N}}$ defined in Eq.~\eqref{eq:reference-product-state}, set $\ket{v}:=J^{\dagger}\ket{\Phi_{r(g),N}}$.
Then $JJ^{\dagger}\ket{\Phi_{r(g),N}}$ is the component of the reference ground state in the embedded hard-core subspace $\mathcal{K}_{A}\otimes\mathcal{H}_{B}$, and $\ket{v}$ is its pullback to the physical spin Hilbert space.
Applying $J^{\dagger}(\,\cdot\,)J$ to the second inequality in Eq.~\eqref{eq:reference-gap} and using Eq.~\eqref{eq:exact-compression}, together with $J^{\dagger}J = \id$ and $J^{\dagger} \ket{\Phi_{r(g),N}} = \ket{v}$, yields
\begin{equation}
	H_{N}(g) \geq Ne_{\mathrm{ref}}(g)\id + \frac{3}{4} \br{\id-\ket{v}\bra{v}}.
	\label{eq:compressed-rank-one-gap}
\end{equation}
Although $\ket{v}$ need not be normalized, its span has dimension at most one.
Every two-dimensional spin subspace therefore contains a normalized vector orthogonal to $\ket{v}$, whose expectation of the right-hand side of Eq.~\eqref{eq:compressed-rank-one-gap} is at least $Ne_{\mathrm{ref}}(g)+3/4$.
The min--max principle then gives
\begin{equation}
	E_{N,1}(g) \geq Ne_{\mathrm{ref}}(g) + \frac{3}{4}.
	\label{eq:first-eigenvalue-lower}
\end{equation}
On the other hand, Proposition~\ref{prop:energy-defect} gives
\begin{equation}
	E_{N,0}(g) = Ne_{\mathrm{ref}}(g) + \delta_{N}(g) \leq Ne_{\mathrm{ref}}(g)+\frac{1}{2}Ng^{4}.
\end{equation}
Hence, $Ng^{4}<3/2$ implies $E_{N,0}(g)<E_{N,1}(g)$, proving uniqueness, and subtraction gives Eq.~\eqref{eq:finite-size-spin-gap}.
For $g=g_{N,\kappa}$, the identity $Ng_{N,\kappa}^{4}=\gamma^{4}N^{3-4\kappa}$ gives Eq.~\eqref{eq:kappa-uniform-gap}.

We next identify the conserved sector containing the unique ground state.
For the normalized ground state $\ket{\Omega}$ of $H_{N}(g)$, with $\ket{\widetilde{\Omega}}:=J\ket{\Omega}$, the exact compression relation gives
$\bra{\widetilde{\Omega}}H_{N}^{\mathrm{ref}}(g)\ket{\widetilde{\Omega}}=E_{N}(g)$.
Applying the second inequality in Eq.~\eqref{eq:reference-gap} to this expectation gives a bound on the overlap between $\ket{\widetilde{\Omega}}$ and the reference product ground state $\ket{\Phi_{r(g),N}}$.
Together with $\delta_{N}(g) = E_{N}(g) - Ne_{\mathrm{ref}}(g) \leq Ng^{4}/2$ from Proposition~\ref{prop:energy-defect} and the condition $Ng^{4} < 3/2$ in Theorem~\ref{thm:uniform-gap}, this yields
\begin{equation}
	1 - \abs{\braket{\Phi_{r(g),N} | \widetilde{\Omega}}}^{2} \leq \frac{4}{3} \delta_{N}(g) < 1.
	\label{eq:reference-overlap}
\end{equation}
Thus, $\ket{\widetilde{\Omega}}$ has nonzero overlap with $\ket{\Phi_{r(g),N}}$.
Since the ground state is unique and $[H_{N}(g),\mathcal{D}]=0$, $\ket{\Omega}$ is an eigenvector of $\mathcal{D} = \mathcal{N}_{A} - \mathcal{N}_{B}$.
Moreover,
$\left(\sum_{b}d_{b}^{\dagger}d_{b}-\mathcal{N}_{B}\right)J=J\mathcal{D}$,
while $\ket{\Phi_{r(g),N}}$ has eigenvalue zero under $\sum_{b}d_{b}^{\dagger}d_{b}-\mathcal{N}_{B}$.
The nonzero overlap therefore forces $\mathcal{D}\ket{\Omega}=0$.
Hence, $\mathcal{N}_{A}\ket{\Omega}=\mathcal{N}_{B}\ket{\Omega}$ and
$\mathcal{N}\ket{\Omega}=2\mathcal{N}_{A}\ket{\Omega}$, so the ground state has even parity.
This completes the proof of Theorem~\ref{thm:uniform-gap}.

\subsection{Excitation Suppression and the Entropy Upper Bound}
\label{subsec:entropy-upper}

We now prove the entropy bounds in Theorem~\ref{thm:finite-size-main}.
Since no ground-state uniqueness is assumed, let $\ket{\Omega}$ denote an arbitrary normalized pure ground state of $H_{N}(g)$ and set $\ket{\widetilde{\Omega}}:=J\ket{\Omega}$.
Taking the expectation of the first inequality in Eq.~\eqref{eq:reference-gap} and using the exact compression relation in Eq.~\eqref{eq:exact-compression-expectation} gives
\begin{equation}
	\sum_{b} \bra{\widetilde{\Omega}} Q_{b}(r(g)) \ket{\widetilde{\Omega}} \leq \frac{4}{3}\delta_{N}(g) \leq \frac{2}{3}Ng^{4} \leq 11Nr(g)^{4}.
	\label{eq:reference-defect-total}
\end{equation}
Here, the second inequality follows from Proposition~\ref{prop:energy-defect}, while the last uses $g=2r(g)/(1-r(g)^{2})$ and $r(g)\leq1/128$.
Thus, the energy defect $\delta_{N}(g)$ controls the total local reference-channel excitation weight $\sum_{b} \bra{\widetilde{\Omega}} Q_{b}(r(g)) \ket{\widetilde{\Omega}}$, whose density is of order $r(g)^{4}$.

We next convert the reference-channel bound in Eq.~\eqref{eq:reference-defect-total} into a bound on the physical excitation numbers.
For the ground state $\ket{\Omega}$, define $m_{A} := \bra{\Omega}\mathcal{N}_{A}\ket{\Omega}$ and $m_{B} := \bra{\Omega}\mathcal{N}_{B}\ket{\Omega}$.
The local reference ground state $\ket{\varphi_{r(g)}}_{b}$ satisfies $\bra{\varphi_{r(g)}} n_{j_{b}} \ket{\varphi_{r(g)}} = r(g)^{2} / (1+r(g)^{2})$.
Since $0 \leq n_{j_{b}} \leq \id$, decomposing a state into the reference-ground-state direction and its orthogonal complement gives
\begin{equation}
	n_{j_{b}} \leq \frac{2r(g)^{2}}{1+r(g)^{2}}\id + 2Q_{b}(r(g)).
	\label{eq:local-number-bound}
\end{equation}
Taking expectations in $\ket{\widetilde{\Omega}}$, summing over $b$, and using Eq.~\eqref{eq:reference-defect-total}, we obtain
\begin{equation}
	m_{B} \leq \frac{2Nr(g)^{2}}{1+r(g)^{2}} + 22Nr(g)^{4}\leq3Nr(g)^{2},
	\label{eq:mB-bound}
\end{equation}
where the last inequality uses $r(g)\leq1/128$.
To obtain a corresponding bound on $m_{A}$, we cannot simply use $m_{A}=m_{B}$: although $\mathcal{D}=\mathcal{N}_{A}-\mathcal{N}_{B}$ is conserved, Theorem~\ref{thm:finite-size-main} does not assume ground-state uniqueness, so an arbitrary pure ground state need not lie in a single $\mathcal{D}$ sector.
We therefore control the imbalance $m_{A}-m_{B}$ directly by defining
\begin{align}
	\Lambda_{b} &:= \sigma_{j_{b}}^{-}-r(g)\sum_{a}(F_{N})_{ab}\sigma_{i_{a}}^{+},
	\label{eq:SOS-Lambda}\\
	\Gamma_{a} &:= \sigma_{i_{a}}^{-}-r(g)\sum_{b}(F_{N})_{ab}\sigma_{j_{b}}^{+}.
	\label{eq:SOS-Gamma}
\end{align}
Using the unitarity of $F_{N}$, these operators satisfy
\begin{align}
	 \sum_{a}\Gamma_{a}^{\dagger} &\Gamma_{a} + \sum_{b}\Lambda_{b}^{\dagger}\Lambda_{b} \notag \\
	 &\,\,\, = (1-r(g)^{2}) \br{H_{N}(g)-Ne_{\mathrm{ref}}(g)\id} ,
\end{align}
and
\begin{align}
	 \sum_{a} \Gamma_{a}^{\dagger} \Gamma_{a} - \sum_{b} \Lambda_{b}^{\dagger} \Lambda_{b}  = (1+r(g)^{2}) (\mathcal{N}_{A}-\mathcal{N}_{B}). 
\end{align} 
Taking expectations of these identities in $\ket{\Omega}$, the sum of the two squared-norm sums equals $(1-r(g)^{2})\delta_{N}(g)$ by Eq.~\eqref{eq:energy-defect-definition}, while the difference of the two squared-norm sums equals $(1+r(g)^{2}) (m_{A}-m_{B})$.
Since both squared-norm sums are nonnegative, applying $|x-y|\leq x+y$ gives
\begin{equation}
	|m_{A}-m_{B}| \leq \frac{1-r(g)^{2}}{1+r(g)^{2}} \delta_{N}(g) \leq \delta_{N}(g).
	\label{eq:imbalance-main}
\end{equation}
Combining this with Eq.~\eqref{eq:mB-bound} and Proposition~\ref{prop:energy-defect}, and using $g=2r(g)/(1-r(g)^{2})$ with $r(g) \leq 1/128$, yields $m_{A} \leq 4Nr(g)^{2}$.
For further details of the operator calculations and numerical estimates used above, see Appendix~\ref{app:upper}.

Finally, we convert the excitation-number bound into an entropy bound.
For each $i_{a}\in A$, let $\rho_{i_{a}}$ be the one-site reduced state of $\ket{\Omega}$ and set $p_{a}:=\bra{\Omega}n_{i_{a}}\ket{\Omega}$.
Subadditivity gives $S_{A}(\ket{\Omega})\leq\sum_{a}S(\rho_{i_{a}})$.
Computational-basis dephasing maps $\rho_{i_{a}}$ to the diagonal state with eigenvalues $1-p_{a}$ and $p_{a}$.
By concavity and unitary invariance of the von Neumann entropy, this dephasing cannot decrease the entropy, and hence
$S(\rho_{i_{a}})\leq h_{2}(p_{a})$.
Using the concavity of $h_{2}$ and $\sum_{a}p_{a}=m_{A}$, we therefore obtain
\begin{align}
	S_{A}(\ket{\Omega}) &\leq \sum_{a}S(\rho_{i_{a}})
	\leq\sum_{a}h_{2}(p_{a}) \notag \\
	&\leq Nh_{2}(m_{A}/N)
	\leq Nh_{2}(4r(g)^{2}).
	\label{eq:entropy-upper-main}
\end{align}
The last inequality follows from $m_{A}/N\leq4r(g)^{2}<1/2$ and the monotonicity of $h_{2}$ on $[0,1/2]$.
Setting $g=g_{N,\kappa}$ proves the upper bound in Theorem~\ref{thm:finite-size-main}.

\subsection{Block Extraction and the Entropy Lower Bound}
\label{subsec:entropy-lower}

We now turn to the entropy lower bound.
Unlike the upper bound, which follows from controlling the physical excitation density, the lower bound exploits the entanglement structure of the reference state through local projections on blocks of channels.
For $G\subseteq\{0,\ldots,N-1\}$ with $|G|=L$, let $A_{G}$ denote the subsystem consisting of the bosonic modes $d_{b}$ with $b\in G$, and let $B_{G}$ denote the subsystem consisting of the corresponding spins $j_{b}$.
Their Hilbert spaces are $\mathcal{F}_{A_{G}}$, the bosonic Fock space of these $L$ modes, and $\mathcal{H}_{B_{G}}:=(\mathbb{C}^{2})^{\otimes L}$, respectively.
Define
\begin{equation}
	\ket{\Phi_{r,G}}:=\bigotimes_{b\in G}\ket{\varphi_{r}}_{b},\quad p_{L}(r):=\frac{Lr^{2}}{(1+r^{2})^{L}}.
	\label{eq:block-reference-and-probability}
\end{equation}
Projecting the reference state $\ket{\Phi_{r,G}}$ in Eq.~\eqref{eq:block-reference-and-probability} onto exactly one boson in $A_{G}$ and one excited spin in $B_{G}$ retains an equal-amplitude superposition of the $L$ terms in which one channel $b \in G$ is in $\ket{1,1}_{b}$ and all other channels are in $\ket{0,0}_{b'}$.
After normalization, the projected state is $\frac{1}{\sqrt{L}}\sum_{b\in G}\left(\ket{1,1}_{b}\otimes\bigotimes_{b'\in G\setminus\{b\}}\ket{0,0}_{b'}\right)$, which is maximally entangled across $A_{G}|B_{G}$ with Schmidt rank $L$.
This outcome occurs with probability $p_{L}(r)$.

To extend the one-pair projection argument to a general block density operator, we use squashed entanglement~\cite{ChristandlWinter}.
For a bipartite density operator $\sigma_{AB}$, it is defined by
\begin{equation}
	E_{\sq}(A:B)_{\sigma_{AB}} := \frac{1}{2}\inf_{\rho_{ABE}}I(A:B|E)_{\rho_{ABE}},
	\label{eq:squashed-definition}
\end{equation}
where the infimum is over all extensions $\rho_{ABE}$ satisfying $\Tr_{E} \rho_{ABE} = \sigma_{AB}$, and $I(A:B|E)_{\rho_{ABE}} := S(\rho_{AE}) + S(\rho_{BE}) - S(\rho_{E}) - S(\rho_{ABE})$ is the conditional mutual information.
The following proposition shows that the one-pair projection gives a lower bound on $E_{\sq}(A_{G}:B_{G})_{\sigma_{G}}$ when the sum of the probabilities that individual channels lie outside their local reference ground states, $\sum_{b\in G}\Tr(\sigma_{G}Q_{b}(r))$, is sufficiently small.

\begin{proposition}
	\label{prop:block-extraction}
	Let $L \geq 4096$, $0 < r \leq 1/128$, and let $\sigma_{G}$ be any density operator on $\mathcal{F}_{A_{G}} \otimes \mathcal{H}_{B_{G}}$ with finite-dimensional local support.
	If
	\begin{equation}
		\epsilon_{G} := \sum_{b\in G} \Tr(\sigma_{G}Q_{b}(r)) \leq \frac{p_{L}(r)^{2}}{100},
		\label{eq:block-good-condition}
	\end{equation}
	then its squashed entanglement satisfies
	\begin{equation}
		E_{\sq}(A_{G}:B_{G})_{\sigma_{G}} \geq \frac{9}{50} p_{L}(r) \log L.
		\label{eq:block-extraction-bound}
	\end{equation}
\end{proposition}

To apply this proposition, let $\ket{\Omega_{N,\kappa}}$ be any normalized pure ground state of the physical Hamiltonian $H_{N,\kappa}$ defined in Eq.~\eqref{eq:fourier-hamiltonian}, and let $\ket{\widetilde{\Omega}_{N,\kappa}}:=J\ket{\Omega_{N,\kappa}}$ denote its embedding into the reference Hilbert space.
For each block $G$, define the reduced density operator $\rho_{G}:=\Tr_{A_{G^{\mathrm{c}}}B_{G^{\mathrm{c}}}}(\ket{\widetilde{\Omega}_{N,\kappa}}\bra{\widetilde{\Omega}_{N,\kappa}})$ on $\mathcal{F}_{A_{G}}\otimes\mathcal{H}_{B_{G}}$, where $G^{\mathrm{c}}:=\{0,\ldots,N-1\}\setminus G$.
Thus, Proposition~\ref{prop:block-extraction} applies to $\rho_{G}$ by taking $\sigma_{G}=\rho_{G}$ whenever the condition in Eq.~\eqref{eq:block-good-condition} is satisfied.
A detailed proof of Proposition~\ref{prop:block-extraction} is provided in Appendix~\ref{app:entanglement-tools}, where we combine local one-pair postselection with properties of squashed entanglement~\cite{ChristandlWinter}, a coherent-information lower bound~\cite{DevetakWinter}, and trace-distance and entropy-continuity estimates~\cite{FuchsVandeGraaf,Audenaert}.
Although the reference Hilbert space is infinite dimensional, the embedded physical ground state contains at most $N$ bosons, and the rotation to the $d_{b}$ modes preserves their total number.
Consequently, every block reduced state $\rho_{G}$ has finite-dimensional support, allowing us to apply the finite-dimensional entropy inequalities.

We next partition the reference channels into blocks and show that sufficiently many of them satisfy the condition of Proposition~\ref{prop:block-extraction}.
Choose $L_{N,\kappa}$ as in Eq.~\eqref{eq:L-def}.
If $L_{N,\kappa}<N$, set $M:=\left\lfloor N/L_{N,\kappa}\right\rfloor$ and partition $ML_{N,\kappa}$ channels into $M$ disjoint blocks $G_{1},\ldots,G_{M}$ of size $L_{N,\kappa}$, discarding the remaining channels.
For each block, define $\epsilon_{G_{\ell}}:=\sum_{b\in G_{\ell}}\Tr(\rho_{G_{\ell}}Q_{b}(r_{N,\kappa}))$.
Since the blocks are disjoint and the operators $Q_{b}(r_{N,\kappa})$ are positive, Eq.~\eqref{eq:reference-defect-total} gives $\sum_{\ell=1}^{M}\epsilon_{G_{\ell}}\leq11Nr_{N,\kappa}^{4}$.
The choice of $L_{N,\kappa}$ also ensures $p_{L_{N,\kappa}}(r_{N,\kappa})>0.30$ and $L_{N,\kappa}\geq8192$, as established in Appendix~\ref{app:global-lower}.
We call a block $G_{\ell}$ good if its reduced density operator $\rho_{G_{\ell}}$ satisfies Eq.~\eqref{eq:block-good-condition} with $\sigma_{G_{\ell}}=\rho_{G_{\ell}}$.
Since every bad block contributes more than $p_{L_{N,\kappa}}(r_{N,\kappa})^{2}/100$ to the sum above, the bound $r_{N,\kappa}^{2}\leq1/16384$ implies that fewer than $0.746Nr_{N,\kappa}^{2}$ blocks are bad, as detailed in Eq.~\eqref{eq:app-bad-count}.
Moreover, $M\geq N/(2L_{N,\kappa})\geq Nr_{N,\kappa}^{2}$, so at least $M/4$ blocks are good.
By Proposition~\ref{prop:block-extraction}, each good block contributes at least $(1/20)\log L_{N,\kappa}$ to the squashed entanglement.
Since the dilation and mode rotation act only on $A$, they preserve the entanglement across $A|B$.
Squashed entanglement equals the entanglement entropy on the embedded pure state, while local discarding and strong superadditivity allow us to sum the contributions from the good blocks.
Since at least $M/4$ blocks are good and each contributes at least $(1/20)\log L_{N,\kappa}$, their total contribution is at least $(M/80)\log L_{N,\kappa}$.
Using $M\geq Nr_{N,\kappa}^{2}$, we obtain
\begin{equation}
	S_{A}(\ket{\Omega_{N,\kappa}})\geq\frac{Nr_{N,\kappa}^{2}}{80}\log L_{N,\kappa}.
	\label{eq:global-lower-main}
\end{equation}

The remaining case $L_{N,\kappa}=N$ consists of a single block $G=\{0,\ldots,N-1\}$, for which $Nr_{N,\kappa}^{2}\leq1/2$.
Using $(1+r_{N,\kappa}^{2})^{N}\leq\ee^{Nr_{N,\kappa}^{2}}$ and $Nr_{N,\kappa}^{2}\leq1/2$, the success probability satisfies $p_{N}(r_{N,\kappa})\geq\ee^{-1/2}Nr_{N,\kappa}^{2}>0.60Nr_{N,\kappa}^{2}$.
On the other hand, Eq.~\eqref{eq:reference-defect-total} gives $\epsilon_{G}\leq11Nr_{N,\kappa}^{4}$.
Consequently, $\epsilon_{G}/p_{N}(r_{N,\kappa})^{2}<11/(0.60^{2}N)<1/100$ for $N\geq4096$, so Proposition~\ref{prop:block-extraction} applies with $\sigma_{G}=\rho_{G}$.
It follows that $S_{A}(\ket{\Omega_{N,\kappa}})\geq(9/50)p_{N}(r_{N,\kappa})\log N>0.108Nr_{N,\kappa}^{2}\log N$, which implies Eq.~\eqref{eq:global-lower-main} since $L_{N,\kappa}=N$.
Appendix~\ref{app:global-lower} provides the detailed estimates for both cases.
Together with Eq.~\eqref{eq:entropy-upper-main}, this completes the proof of Theorem~\ref{thm:finite-size-main}.

\section{Weak-Coupling Analysis}
\label{sec:weak-coupling}
 
We now derive explicit weak-coupling formulas that explain the pairing structure and provide finite-size comparison curves.
For the ground-state entanglement, we use quadratic Holstein--Primakoff theory, whereas for the physical gap we use perturbation theory in the original spin Hilbert space.

\subsection{Quadratic Bosonic Hamiltonian and Ground State}
\label{subsec:HP-model}

The excitation-number bounds in Eqs.~\eqref{eq:mB-bound} and \eqref{eq:imbalance-main} give $m_{A}/N \leq 4r(g)^{2}$ and $m_{B}/N \leq 3r(g)^{2}$, with $r(g)^{2}=\cO(g^{2})$ at weak coupling.
For $g=g_{N,\kappa}$, these upper bounds vanish as $N \to \infty$ when $\kappa>1/2$, while at $\kappa=1/2$ they remain small for sufficiently small $\gamma$.
This motivates a Holstein--Primakoff expansion around the polarized vacuum~\cite{HolsteinPrimakoff1940}.

We use the bosonic modes $a_{a}$ introduced in Sec.~\ref{subsec:exact-dilation} and introduce additional bosonic modes $b_{b}$ for subsystem $B$.
At quadratic order, we use
$n_{i_{a}}\approx a_{a}^{\dagger}a_{a}$, $\sigma_{i_{a}}^{+}\approx a_{a}^{\dagger}$, and $\sigma_{i_{a}}^{-}\approx a_{a}$ on $A$, together with
$n_{j_{b}}\approx b_{b}^{\dagger}b_{b}$, $\sigma_{j_{b}}^{+}\approx b_{b}^{\dagger}$, and $\sigma_{j_{b}}^{-}\approx b_{b}$ on $B$.
Applying these replacements to $H_{N}(g)=\mathcal{N}-g(P_{N}^{\dagger}+P_{N})$, defined in Eq.~\eqref{eq:fourier-hamiltonian-g}, with the pairing operators defined in Eq.~\eqref{eq:P-def}, gives
\begin{align}
	H_{\mathrm{HP}}^{(2)}(g) &=\sum_{a} a_{a}^{\dagger} a_{a}+\sum_{b} b_{b}^{\dagger} b_{b} \notag \\
	&\quad -g\sum_{a,b}\Bigl( (F_{N})_{ab}a_{a}^{\dagger} b_{b}^{\dagger} +\overline{(F_{N})_{ab}}a_{a}b_{b}\Bigr).
	\label{eq:HP-site-hamiltonian}
\end{align}
In contrast to Eq.~\eqref{eq:reference-hamiltonian}, this approximation replaces the spins on both sides by unconstrained bosons.
After the $A$-local rotation in Eq.~\eqref{eq:rotated-bosons}, Eq.~\eqref{eq:HP-site-hamiltonian} becomes a sum of identical two-mode pairing Hamiltonians,
\begin{equation}
	H_{\mathrm{HP}}^{(2)}(g) =\sum_{b=0}^{N-1}\Bigl[ d_{b}^{\dagger} d_{b}+b_{b}^{\dagger} b_{b} -g(d_{b}^{\dagger} b_{b}^{\dagger}+d_{b}b_{b})\Bigr].
	\label{eq:HP}
\end{equation}
For $0\leq g<1$, define
\begin{equation}
	\lambda(g):=\frac{g}{1+\sqrt{1-g^{2}}}.
	\label{eq:HP-parameters}
\end{equation}
A Bogoliubov transformation diagonalizes each block in Eq.~\eqref{eq:HP}, yielding the Gaussian ground state $\ket{\Omega_{\mathrm{HP}}}$ of $H_{\mathrm{HP}}^{(2)}(g)$ in the form
\begin{equation}
	\ket{\Omega_{\mathrm{HP}}} =\bigotimes_{b=0}^{N-1} \left[\sqrt{1-\lambda(g)^{2}} \sum_{m=0}^{\infty} \lambda(g)^{m} \ket{m}_{d_{b}}\ket{m}_{b_{b}}\right].
	\label{eq:HP-ground-state}
\end{equation}
The detailed Bogoliubov transformation leading to Eq.~\eqref{eq:HP-ground-state} is given in Appendix~\ref{app:weak-coupling}.

\subsection{Entanglement Entropy of the Quadratic Ground State}
\label{subsec:HP-entropy}

We now evaluate the bipartite entanglement entropy of the quadratic ground state in Eq.~\eqref{eq:HP-ground-state}.
Because the Holstein--Primakoff approximation replaces the spins on both subsystems by bosons, this state is a pure Gaussian state.
Moreover, Eq.~\eqref{eq:HP-ground-state} factorizes into $N$ identical two-mode squeezed states after the $A$-local rotation in Eq.~\eqref{eq:rotated-bosons}.
Since this rotation acts only within $A$, it preserves the $A|B$ entanglement.
For each two-mode squeezed pair in Eq.~\eqref{eq:HP-ground-state}, tracing out one mode gives a reduced state that is diagonal in the occupation basis, with probability $(1-\lambda(g)^{2})\lambda(g)^{2m}$ for the $m$-boson state.
The corresponding mean occupation of the geometric distribution is
\begin{equation}
	\nu(g):=\frac{\lambda(g)^{2}}{1-\lambda(g)^{2}} =\frac{1}{2}\br{\frac{1}{\sqrt{1-g^{2}}}-1}.
	\label{eq:HP-occupation}
\end{equation}
Summing the entropy over the $N$ independent pairs gives
\begin{align}
	S_{A}^{\mathrm{HP}}(g) &= N\big[(\nu(g)+1)\log(\nu(g)+1) \notag \\ &\qquad \qquad \qquad \qquad -\nu(g)\log\nu(g) \big].
	\label{eq:HP-entropy-exact}
\end{align}
This is the standard Gaussian entropy expressed through the reduced occupation number~\cite{WeedbrookEtAl2012}.

Using $\nu(g) = g^{2} / 4 + 3g^{4} / 16 + \cO(g^{6})$ gives
\begin{equation}
	S_{A}^{\mathrm{HP}}(g)=\frac{Ng^{2}}{4}\brr{1-\log\br{\frac{g^{2}}{4}}}+\cO\br{Ng^{4}|\log g|}.
	\label{eq:HP-entropy-leading}
\end{equation}
For fixed $\gamma>0$ and $1/2<\kappa\leq1$, substituting $g_{N,\kappa}=\gamma N^{\frac{1}{2}-\kappa}$ into Eq.~\eqref{eq:HP-entropy-leading} gives
\begin{align}
	S_{A}^{\mathrm{HP}}(g_{N,\kappa})
	&=\frac{\gamma^{2}}{4}N^{2-2\kappa}(2\kappa-1)\log N \notag \\
	&\quad+\frac{\gamma^{2}}{4}N^{2-2\kappa}
	\brr{1-\log\br{\frac{\gamma^{2}}{4}}} \notag \\
	&\quad+\cO_{\gamma,\kappa}\br{\gamma^{4}N^{3-4\kappa}\log N}.
	\label{eq:HP-entropy-N}
\end{align}
In particular, at the natural Kac normalization $\kappa = 1$,
\begin{align}
	S_{A}^{\mathrm{HP}}(g_{N,\kappa})\big|_{\kappa=1}
	&=\frac{\gamma^{2}}{4}\brr{1+\log\br{\frac{4N}{\gamma^{2}}}} \notag \\
	&\qquad \quad + \cO_{\gamma}\br{\frac{\gamma^{4}\log N}{N}}.
	\label{eq:HP-entropy-Kac}
\end{align}
For $1/2 <\kappa<1$, the leading growth is $N^{2-2\kappa}\log N$.
At $\kappa = 1/2$, $g_{N,\kappa}=\gamma$ is independent of $N$, and the exact quadratic formula becomes
\begin{equation}
	S_{A}^{\mathrm{HP}}(g_{N,\kappa})=N s_{\mathrm{HP}}(\gamma),
	\label{eq:HP-volume-law}
\end{equation}
where $s_{\mathrm{HP}}(g) := (\nu(g)+1)\log(\nu(g)+1) - \nu(g)\log\nu(g)$.
It satisfies $s_{\mathrm{HP}}(g) > 0$ for $0 < g < 1$.
These quadratic formulas reproduce the entanglement scaling rigorously established for every pure ground state of the original spin Hamiltonian in Theorem~\ref{thm:finite-size-main}.

\subsection{The Physical Spin Gap from Degenerate Perturbation Theory}
\label{subsec:spin-gap-perturbation}

We now turn to the physical spectral gap of the original spin Hamiltonian.
To determine its leading weak-coupling correction, we apply perturbation theory directly to $H_{N}(g)$ at fixed finite $N$.
At $g=0$, the vacuum is the unique ground state, the one-spin manifold has energy $1$ and dimension $2N$, and all remaining excitation sectors start at energy $2$ or higher.
For each fixed $N$, the spectral clusters remain separated and perturbation theory applies for sufficiently small $|g|$~\cite{Kato1995}.

We first compute the ground-state energy.
Since $P_{N} \ket{\mathbf{0}}=0$ and $\norm{P_{N}^{\dagger} \ket{\mathbf{0}}}^{2}=N$, the first-order correction to the vacuum energy vanishes, while the second-order correction comes from intermediate states with two excitations and energy denominator $-2$.
Thus,
\begin{equation}
	E_{N,0}(g) = -\frac{N}{2}g^{2} + \cO_{N}(g^{4}).
	\label{eq:spin-E0-perturbative}
\end{equation}
Odd-order corrections vanish because the pairing terms change the total excitation number by two.

We next compute the energy of the one-spin excited manifold, starting from the one-$A$-spin state $\ket{c}_{A} := \sigma_{i_{c}}^{+}\ket{\mathbf{0}}$.
Pair annihilation vanishes on this state, while pair creation gives
\begin{align}
	P_{N}^{\dagger}\ket{c}_{A} = \sum_{\substack{a\neq c\\ b=0,\ldots,N-1}}(F_{N})_{ab} \sigma_{i_{a}}^{+}\sigma_{j_{b}}^{+}\ket{c}_{A}.
	\label{eq:blocked-pair-state}
\end{align}
The term with $a=c$ is absent because the physical spin is already excited.
Unitarity therefore gives the diagonal matrix element
\begin{equation}
	\bra{c}_{A} P_{N}P_{N}^{\dagger} \ket{c}_{A} = \sum_{a\neq c,b}|(F_{N})_{ab}|^{2} = N-1.
	\label{eq:blocked-norm}
\end{equation}
This value is independent of $c$.
For two distinct one-$A$-spin states $\ket{c}_{A}$ and $\ket{c'}_{A}$, the only intermediate states reached from both have $A$-spin excitations at $c$ and $c'$ and a single $B$-spin excitation at $b$.
Hence, for $c\neq c'$,
\begin{equation}
	\bra{c'}_{A}P_{N}P_{N}^{\dagger}\ket{c}_{A} = \sum_{b}(F_{N})_{c'b}\overline{(F_{N})_{cb}} = 0
\end{equation}
by orthogonality of distinct rows of $F_{N}$.
Thus, $P_{N}P_{N}^{\dagger}$ acts as $(N-1)\id$ on the one-$A$-spin subspace.
The same argument for the one-$B$-spin subspace uses column orthogonality and gives the same eigenvalue $N-1$.
Moreover, the one-$A$-spin and one-$B$-spin subspaces have different $\mathcal{D}$ eigenvalues and therefore cannot mix.
If $\mathscr{P}_{1}$ denotes the projector onto the full one-spin manifold, then
\begin{equation}
	\mathscr{P}_{1} P_{N}P_{N}^{\dagger} \mathscr{P}_{1} = (N-1)\mathscr{P}_{1}.
	\label{eq:PPdagger-one}
\end{equation}

The intermediate states now have three excitations, so the denominator is again $1-3=-2$.
As in the ground-state expansion, all odd-order corrections vanish because each application of the pairing perturbation changes the total excitation number by two.
For fixed $N$ and sufficiently small $|g|$, the eigenvalues emerging from the one-spin manifold form the first excited cluster.
Its lowest eigenvalue therefore satisfies
\begin{equation}
	E_{N,1}(g) = 1 - \frac{N-1}{2}g^{2} + \cO_{N}(g^{4}).
	\label{eq:spin-E1-perturbative}
\end{equation}
Subtracting Eq.~\eqref{eq:spin-E0-perturbative} from Eq.~\eqref{eq:spin-E1-perturbative} gives
\begin{equation}
	\Delta_{N}(g) = 1 + \frac{1}{2}g^{2} + \cO_{N}(g^{4}).
	\label{eq:spin-gap-perturbative}
\end{equation}
The positive second-order correction to the gap has a direct hard-core origin.
For the vacuum, the total virtual pair-creation weight is $N$, whereas for a one-spin state it is reduced to $N-1$, because pair creation involving the already excited site is forbidden.
Consequently, the vacuum energy receives a correction of $-Ng^{2}/2$, while the one-spin energy receives a correction of only $-(N-1)g^{2}/2$.
Their difference therefore increases the physical gap by $g^{2}/2$ at second order.

Substitution of $g = g_{N,\kappa} = \gamma N^{\frac{1}{2}-\kappa}$ gives the formal fixed-size expression
\begin{equation}
	\Delta_{N}(g_{N,\kappa}) =1+\frac{\gamma^{2}}{2}N^{1-2\kappa}+R_{N}(g_{N,\kappa}),
	\label{eq:gap-N-formal}
\end{equation}
where $R_{N}(g)=\cO_{N}(g^{4})$ as $g\to0$ for fixed $N$.

\section{Numerical Results}
\label{sec:numerics}

We compare numerical results for the Fourier-paired all-to-all bipartite qubit Hamiltonian $H_{N,\kappa}$ in Eq.~\eqref{eq:fourier-hamiltonian} with the analytic formulas derived above, focusing on the physical spin gap and the ground-state entanglement across $A|B$.
The numerical comparisons use DMRG~\cite{white1992density,white1993density,schollwock2011density} with the ITensor library~\cite{Fishman2022ITensor}.
For the spectral gap, we compute the ground-state and first-excited-state energies in both the even- and odd-parity sectors, collect these four energies, and define the gap as the difference between the two lowest energies.
In each parity sector, the ground-state optimization uses $5$ sweeps, whereas the first-excited-state optimization uses $15$ sweeps.
Throughout this section, the coupling is fixed at $\gamma = 0.01$, the system size is expressed in terms of the total number of spins $n = 2N$, and the effective coupling is $g_{N,\kappa} = 0.01 \,N^{\frac{1}{2}-\kappa}$.
This value of $\gamma$ satisfies the condition $\gamma \leq 1/64$ assumed in the main theorems, although the finite system sizes used numerically do not satisfy the requirement $N \geq 4096$ in Theorem~\ref{thm:finite-size-main}.
The numerical comparison is therefore a finite-size illustration rather than a direct test of the finite-size entropy theorem in its stated size range.

\begin{figure}[!tbp]
	\centering
	\includegraphics[width=0.45\textwidth]{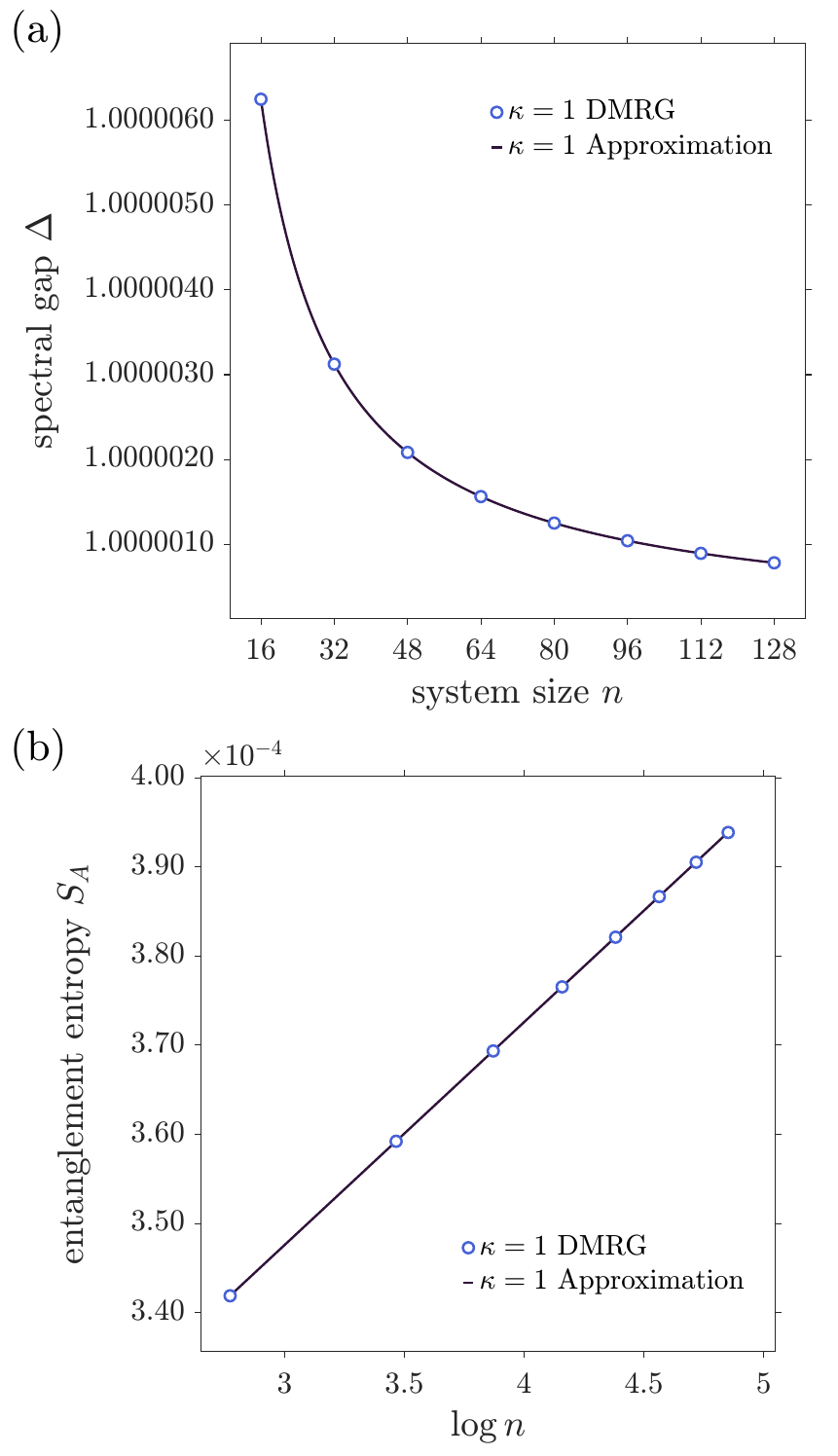}
	\caption{Finite-size comparison for the Fourier-paired spin model in Eq.~\eqref{eq:fourier-hamiltonian} at the natural Kac normalization $\kappa=1$ and $\gamma=0.01$.
		Panel (a) compares the physical spin gap with the second-order spin perturbation curve $1+ g_{N,\kappa}^{2}/2$ from Eq.~\eqref{eq:spin-gap-perturbative}.
		Panel (b) compares the bipartite entropy with the quadratic Holstein--Primakoff formula in Eq.~\eqref{eq:HP-entropy-exact}.
		The circles denote Fourier-model DMRG data for the system sizes considered, up to $n=128$ ($N=64$), with a maximum bond dimension of $2000$.
		The solid lines represent the corresponding analytic comparison curves, evaluated with $N=n/2$.}
	\label{fig:Fourier-kappa1}
\end{figure}

We first consider $\kappa=1$.
Figure~\ref{fig:Fourier-kappa1} is organized into a physical-gap panel and an entropy panel, using system sizes up to $n=128$ and a maximum matrix product state bond dimension of approximately $2000$.
For the gap, the weak-coupling comparison curve is $1 + \gamma^{2}/(2N)$ from Eq.~\eqref{eq:spin-gap-perturbative}.
For the entropy, the comparison curve is Eq.~\eqref{eq:HP-entropy-exact}, whose large-$N$ form is logarithmic at $\kappa=1$.
The DMRG data closely follow both analytic comparison curves over the system sizes considered.
In particular, the entropy panel shows an approximately linear dependence on $\log n$, providing numerical support for the predicted logarithmic scaling.

\begin{figure}[!tbp]
	\centering
	\includegraphics[width=0.48\textwidth]{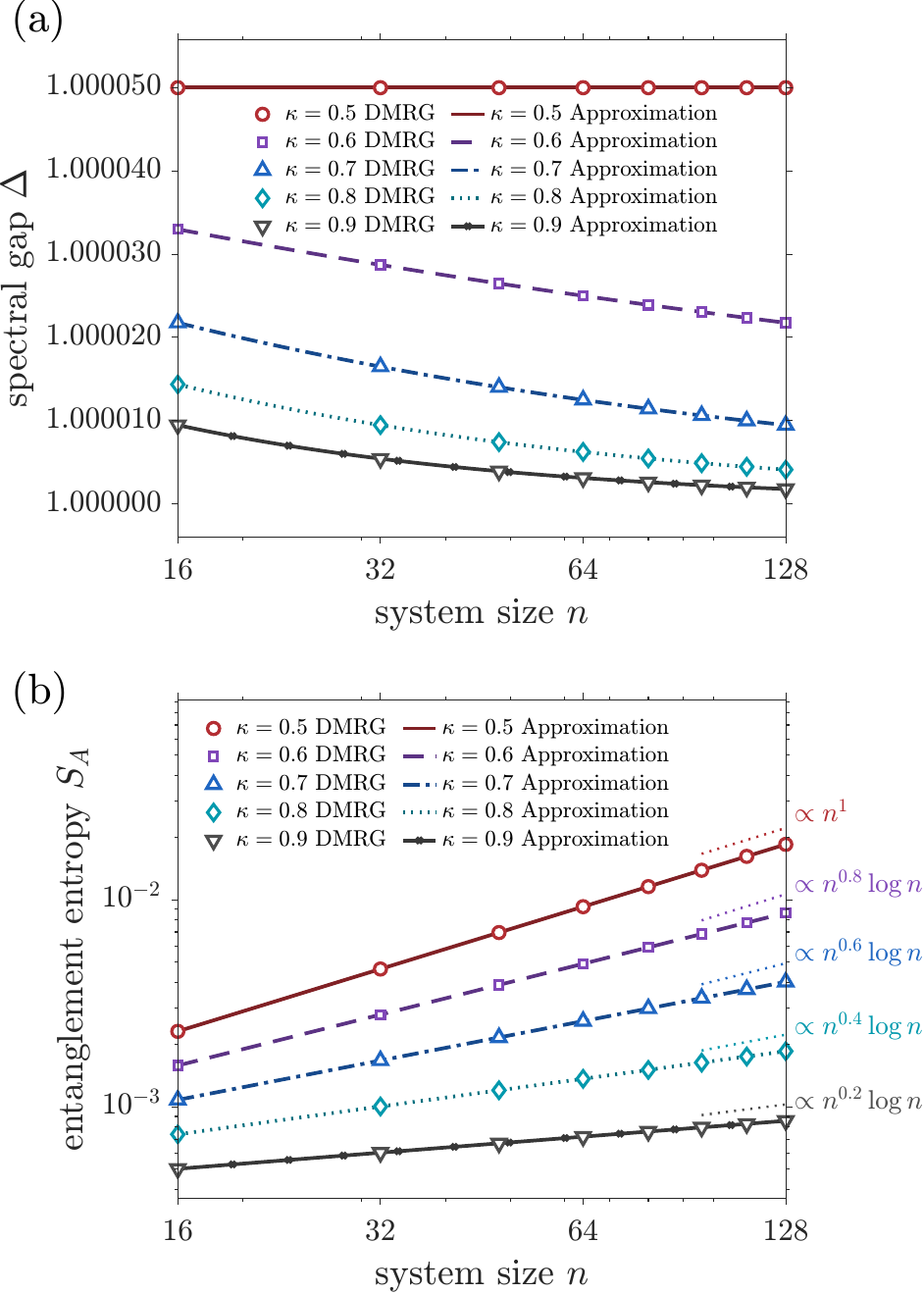}
	\caption{Fourier-paired model in Eq.~\eqref{eq:fourier-hamiltonian} for $\kappa=0.5,0.6,0.7,0.8,0.9$ at fixed $\gamma=0.01$.
		Panel (a) shows the physical spin gap, and panel (b) shows the bipartite entanglement entropy across $A|B$.
		The markers represent Fourier-model DMRG data, while the corresponding lines show the analytic comparison curves evaluated with $g_{N,\kappa}=0.01\,N^{\frac{1}{2}-\kappa}$ and $N=n/2$; the colors and marker or line styles distinguish the different values of $\kappa$.
		For panel (a), the comparison curve is the second-order spin perturbation formula $1+g_{N,\kappa}^{2}/2$ from Eq.~\eqref{eq:spin-gap-perturbative}.
		For panel (b), the comparison curve is the quadratic Holstein--Primakoff entropy formula in Eq.~\eqref{eq:HP-entropy-exact}.
		The thin dotted guide segments in panel (b) illustrate the asymptotic entanglement scalings $S_{A} = \Theta(n)$ at $\kappa = 1/2$ and $S_{A} = \Theta(n^{2 - 2\kappa} \log n)$ for $1/2 < \kappa < 1$.
		The calculations use system sizes up to $n=128$ ($N=64$); the maximum bond dimension is approximately $10000$ for the gap calculations in panel (a) and approximately $2000$ for the entropy calculations in panel (b).}
	\label{fig:Fourier-general}
\end{figure}

We next compare both the gap and the entropy for several values of $\kappa$ in Fig.~\ref{fig:Fourier-general}.
Panel (a) collects the physical spin-gap data for $\kappa=0.5,0.6,0.7,0.8,0.9$, and panel (b) collects the corresponding entropy data for the same values of $\kappa$.
The gap calculations use a maximum bond dimension of approximately $10000$, while the entropy calculations use a maximum bond dimension of approximately $2000$.
For the gap, the comparison is again made with the spin perturbative expression $1+g_{N,\kappa}^{2}/2$.
For the entropy, the main analytic comparison uses the full quadratic formula in Eq.~\eqref{eq:HP-entropy-exact}.
Panel (b) additionally shows thin dotted guides proportional to $n^{2-2\kappa}\log n$ for $1/2 < \kappa<1$, together with a linear guide proportional to $n$ at the endpoint $\kappa=1/2$.  
Across the values of $\kappa$ considered, the DMRG data show good agreement with the corresponding analytic comparison curves.
For the entropy, the numerical growth is also consistent with the leading scaling forms indicated by the dotted guides.

\section{Discussion}
\label{sec:discussion}

In this work, we studied an explicit Fourier-paired all-to-all bipartite qubit Hamiltonian and established its ground-state entanglement scaling as a function of the interaction normalization.
The central result is that the logarithmic entanglement scaling at the natural Kac normalization $\kappa=1$ is immediately lost once the normalization exponent is decreased below $\kappa=1$.
For every $1/2<\kappa<1$, every pure ground state has the parametrically stronger but still subextensive entropy $\Theta(n^{2-2\kappa}\log n)$, while the endpoint $\kappa=1/2$ exhibits a volume law.
At $\kappa=1$, the ground state is nondegenerate and uniformly gapped, and its logarithmic entanglement scaling matches the optimal order known for Kac-normalized fully connected systems~\cite{KimKuwahara2024,KimKuwahara2026Collectivity}.

For $3/4\leq\kappa<1$, these stronger entropy scalings occur in a unique ground state separated from the rest of the spectrum by a system-size-independent gap.
For $1/2\leq\kappa<3/4$, a regime beyond the range covered by Theorem~\ref{thm:uniform-gap}, the weak-coupling analysis and DMRG calculations provide complementary evidence for ground-state uniqueness and a nonclosing physical gap.
Thus, changing the interaction normalization alone generates a controlled hierarchy from logarithmic entanglement at the natural Kac normalization through progressively stronger subextensive scaling and ultimately to a volume law, showing that even an arbitrarily weak relaxation of the natural Kac normalization precludes a universal logarithmic entanglement bound.
The construction also separates the growth of the local interaction scale from that of the ground-state energy density: for $\kappa<1$ the incident interaction norm grows with system size, whereas the ground-state energy density still vanishes for every $\kappa>1/2$.

Several open problems remain in extending and sharpening these results.
First, rigorously establishing a uniform positive spectral gap for the physical spin Hamiltonian in the regime $1/2\leq\kappa<3/4$ remains an open problem.
Second, the optimal universal entanglement upper bound for nondegenerate uniformly gapped fully connected Hamiltonians with bounded on-site terms and pairwise interactions of strength $\cO(n^{-\kappa})$ remains unknown for $1/2<\kappa<1$.
In the rigorously gapped regime $3/4\leq\kappa<1$, our construction achieves $\Theta(n^{2-2\kappa}\log n)$, whereas rescaling the existing general bound~\cite{KimKuwahara2024} gives the weaker upper bound $\cO(n^{3-3\kappa}\log n)$.
This discrepancy corresponds to the gap between the quadratic lower bound in Eq.~\eqref{eq:kac-gap-lower} and the cubic spectral-gap dependence of the existing general upper bound.
Whether the quadratic dependence can be achieved as a universal upper bound remains open.
Resolving these questions would clarify the optimal constraints on ground-state entanglement imposed by interaction normalization and the spectral gap.

\section*{Acknowledgements}

Both authors acknowledge support from the RIKEN Hakubi Project.
D.K. was supported by the RIKEN Special Postdoctoral Researcher Program and JSPS KAKENHI Grant Number JP26K17060.
T.K. was supported by the Japan Science and Technology Agency through its Exploratory Research for Advanced Technology program (JST ERATO Grant Number JPMJER2302) and by JSPS KAKENHI Grant Numbers JP23K25796, JP25K24674, JP26K00019, JP26K00020, and JP26H02015.
The authors used OpenAI's GPT-6 model to assist in developing and refining proof ideas and improving the clarity and presentation of the manuscript. All results, proofs, calculations, and references were independently verified by the authors, who take full responsibility for the content of the manuscript.

\bibliography{Fourier_all_to_all}

\appendix

\section{Exact Dilation and the Reference Spectrum}
\label{app:dilation}

In this appendix, we provide the detailed derivations of the exact compression relation in Eq.~\eqref{eq:exact-compression} and the reference spectral bounds in Eq.~\eqref{eq:reference-gap}.

\subsection{Hard-Core Embedding and Exact Compression}

Recall from Sec.~\ref{subsec:exact-dilation} that $\mathcal{K}_{A}\subset\mathcal{F}_{A}$ is the subspace spanned by bosonic occupation vectors with occupation zero or one in each original mode $a_{a}$.
For a subset $R\subseteq\{0,\ldots,N-1\}$, the vector $\prod_{a\in R}a_{a}^{\dagger}\ket{0}_{\mathcal{F}_{A}}$ has norm one, and different subsets give orthogonal vectors.
Thus, Eq.~\eqref{eq:hardcore-isometry} defines an isometry from the $A$ spin Hilbert space onto $\mathcal{K}_{A}$.
An annihilation operator removes an occupied mode with coefficient one.
A creation operator adds an unoccupied mode with coefficient one, while creation in an already occupied mode leaves $\mathcal{K}_{A}$ and is removed by the final compression.
It follows that
\begin{align}
	J_{A}^{\dagger} a_{a}J_{A} &= \sigma_{i_{a}}^{-},
	\label{eq:app-compression-annihilation}\\
	J_{A}^{\dagger} a_{a}^{\dagger} J_{A} &= \sigma_{i_{a}}^{+},
	\label{eq:app-compression-creation}\\
	J_{A}^{\dagger} a_{a}^{\dagger} a_{a}J_{A} &= n_{i_{a}}.
	\label{eq:app-compression-number}
\end{align}

For the modes in Eq.~\eqref{eq:rotated-bosons}, unitarity gives $[d_{b},d_{b'}^{\dagger}]=\sum_{a}\overline{(F_{N})_{ab}}(F_{N})_{ab'}=\delta_{bb'}$ and the identities
\begin{align}
	\sum_{b}d_{b}^{\dagger}d_{b}
	&=\sum_{a}a_{a}^{\dagger}a_{a},
	\label{eq:app-mode-number-identity}\\
	\sum_{b}d_{b}^{\dagger}\sigma_{j_{b}}^{+}
	&=\sum_{a,b}(F_{N})_{ab}a_{a}^{\dagger}\sigma_{j_{b}}^{+}.
	\label{eq:app-mode-pair-identity}
\end{align}
Compressing the last two identities with $J=J_{A}\otimes\id_{B}$ proves Eq.~\eqref{eq:exact-compression}.
Since the reference Hamiltonian has ground energy $Ne_{\mathrm{ref}}(g)$, Eq.~\eqref{eq:exact-compression-expectation} gives
\begin{equation}
	E_{N}(g)\geq Ne_{\mathrm{ref}}(g).
	\label{eq:app-dilation-energy-lower}
\end{equation}
The embedding is local with respect to $A|B$ and does not change the Schmidt coefficients.
The change from $a$ to $d$ modes is a number-preserving unitary transformation within the same bosonic subsystem and is also local with respect to that bipartition.

\subsection{Local Boson--Spin Blocks}

Omit the channel label and consider
\begin{equation}
	h(g)=d^{\dagger} d+n-g(d^{\dagger}\sigma^{+}+d\sigma^{-}).
	\label{eq:app-local-reference}
\end{equation}
It commutes with $d^{\dagger} d-n$.
The eigenspace of $d^{\dagger}d-n$ with eigenvalue $-1$ is spanned by $\ket{0,1}$, which has energy $1$.
For every integer $\ell\geq0$, the eigenspace with eigenvalue $\ell$ is spanned by $\ket{\ell,0}$ and $\ket{\ell+1,1}$.
In this ordered basis,
\begin{equation}
	h(g)\big|_{\ell}=
	\begin{pmatrix}
		\ell&-g\sqrt{\ell+1}\\
		-g\sqrt{\ell+1}&\ell+2
	\end{pmatrix},
	\label{eq:local-reference-matrix}
\end{equation}
with eigenvalues
\begin{equation}
	e_{\ell,\pm}(g)=\ell+1\pm\sqrt{1+g^{2}(\ell+1)}.
	\label{eq:local-reference-eigenvalues}
\end{equation}
Since $e_{\ell,-}(g)\to\infty$ as $\ell\to\infty$, the orthogonal direct sum of these finite-dimensional blocks defines a self-adjoint operator $h(g)$ that is bounded below.  

\begin{lemma}
	\label{lem:reference-spectrum}
	For $0\leq g\leq1/4$, the unique ground state of $h(g)$ is $\ket{\varphi_{r(g)}}$ in Eq.~\eqref{eq:reference-local-ground}, with ground energy $e_{\mathrm{ref}}(g)=1-\sqrt{1+g^{2}}$.
	Moreover,
	\begin{equation}
		h(g)-e_{\mathrm{ref}}(g)\id \geq\frac{3}{4}\br{\id-\ket{\varphi_{r(g)}}\bra{\varphi_{r(g)}}}.
		\label{eq:app-local-reference-gap}
	\end{equation}
\end{lemma}

\begin{proof}
	The lower eigenvalue for $\ell=0$ is $1-\sqrt{1+g^{2}}=e_{\mathrm{ref}}(g)$.
	The corresponding eigenvector equation gives the ratio of the $\ket{1,1}$ and $\ket{0,0}$ amplitudes as $r(g)=g/(1+\sqrt{1+g^{2}})$.
	For $\ell\geq1$, define $f(x):=x-\sqrt{1+g^{2}x}$ for $x\geq1$.
	Since $0 \leq g \leq 1/4$, we have
	\begin{align}
		f'(x) =1-\frac{g^{2}}{2\sqrt{1+g^{2}x}} \geq 1-\frac{g^{2}}{2}\geq\frac{31}{32}.
	\end{align}
	Consequently,
	\begin{align}
		e_{\ell,-}(g)-e_{\mathrm{ref}}(g) = f(\ell+1)-f(1) \geq\frac{31}{32}\ell.
	\end{align}
	The isolated vector $\ket{0,1}$ lies at energy distance $\sqrt{1+g^{2}}\geq1$, and the upper $\ell=0$ eigenvalue lies at distance $2\sqrt{1+g^{2}}\geq2$.
	All other upper eigenvalues are at least as large as their corresponding lower eigenvalues.
	Thus, the local ground state is unique and its gap is at least $31/32$, which implies the weaker bound in Eq.~\eqref{eq:app-local-reference-gap}.
\end{proof}

Since the $N$ reference terms act on distinct mode--spin tensor factors, their unique ground state is $\ket{\Phi_{r(g),N}}$ and their ground energies add.
Summing Eq.~\eqref{eq:app-local-reference-gap} gives
\begin{equation}
	H_{N}^{\mathrm{ref}}(g)-Ne_{\mathrm{ref}}(g)\id \geq\frac{3}{4}\sum_{b}Q_{b}(r(g)).
\end{equation}
The projectors $Q_{b}(r(g))$ commute.
It follows that
\begin{align}
	\sum_{b}Q_{b}(r(g)) &\geq \id-\prod_{b}(\id-Q_{b}(r(g))) \notag \\
	&= \id-\ket{\Phi_{r(g),N}}\bra{\Phi_{r(g),N}}.
\end{align}
This completes the proof of Eq.~\eqref{eq:reference-gap}.

\section{Variational Bound on the Reference-Energy Defect}
\label{app:variational}

In this appendix, we provide the detailed variational proof of Proposition~\ref{prop:energy-defect}.
Recall the Fourier pairing operators $P_{N}^{\dagger}$ and $P_{N}$ from Eq.~\eqref{eq:P-def}.
Both operators commute with $\mathcal{D}=\mathcal{N}_{A}-\mathcal{N}_{B}$, as stated in Eq.~\eqref{eq:conserved-quantities}.
On the equal-excitation subspace $\mathcal{H}_{\mathrm{eq}}:=\ker\mathcal{D}$, we write $T:=\mathcal{N}_{A}|_{\mathcal{H}_{\mathrm{eq}}}=\mathcal{N}_{B}|_{\mathcal{H}_{\mathrm{eq}}}$, as in Sec.~\ref{subsec:energy-defect}.
All operators in this section act on the original finite-dimensional spin Hilbert space.

\subsection{A Positive One-Particle Matrix Estimate}

For either subsystem $A$ or $B$, let $p,q$ index its $N$ sites.
If an $N\times N$ matrix $K=(K_{pq})$ satisfies $0\leq K\leq\id_{N}$, then
\begin{equation}
	0\leq\sum_{p,q}K_{pq}\sigma_{p}^{+}\sigma_{q}^{-}\leq\sum_{p}n_{p}.
	\label{eq:positive-matrix-spin-bound}
\end{equation}
To prove the first inequality in Eq.~\eqref{eq:positive-matrix-spin-bound}, factor $K=R^{\dagger}R$.
Then
\begin{align}
	0&\leq \sum_{\ell}\left(\sum_{p}R_{\ell p}\sigma_{p}^{-}\right)^{\dagger}
	\left(\sum_{q}R_{\ell q}\sigma_{q}^{-}\right) \notag \\
	&= \sum_{p,q}K_{pq}\sigma_{p}^{+}\sigma_{q}^{-}.
\end{align}
To prove the second inequality in Eq.~\eqref{eq:positive-matrix-spin-bound}, apply the same argument to $\id_{N}-K\geq0$, which gives
\begin{align}
	0 &\leq \sum_{p,q}(\delta_{pq}-K_{pq})\sigma_{p}^{+}\sigma_{q}^{-}  = \sum_{p}n_{p} - \sum_{p,q} K_{pq} \sigma_{p}^{+} \sigma_{q}^{-}.
\end{align}

\subsection{The Pair-Commutator Bound}

Define
\begin{align}
	q_{a}&:=\sum_{b}\overline{(F_{N})_{ab}}\sigma_{j_{b}}^{-},
	\label{eq:app-q-operator}\\
	e_{b}&:=\sum_{a}\overline{(F_{N})_{ab}}\sigma_{i_{a}}^{-}.
	\label{eq:app-e-operator}
\end{align}
Then $P_{N}=\sum_{a}\sigma_{i_{a}}^{-}q_{a}=\sum_{b}e_{b}\sigma_{j_{b}}^{-}$.
The spin commutators are $[\sigma_{i_{a}}^{-},\sigma_{i_{a'}}^{+}]=\delta_{aa'}(\id-2n_{i_{a}})$ and $[q_{a},q_{a'}^{\dagger}]=\delta_{aa'}\id-2\sum_{b}\overline{(F_{N})_{ab}}(F_{N})_{a'b}n_{j_{b}}$.
Expanding $[P_{N},P_{N}^{\dagger}]$ and using unitarity yields
\begin{align}
	[P_{N},P_{N}^{\dagger}]
	&=N\id-\mathcal{N}_{B}+\mathcal{N}_{A} - 2\mathcal{X} - 2\mathcal{Y},
	\label{eq:app-commutator-decomposition}\\
	\mathcal{X}&:=\sum_{a}n_{i_{a}}q_{a}q_{a}^{\dagger},
	\label{eq:app-X-definition}\\
	\mathcal{Y}&:=\sum_{b}n_{j_{b}}e_{b}^{\dagger}e_{b}.
	\label{eq:app-Y-definition}
\end{align}
Both $\mathcal{X}$ and $\mathcal{Y}$ are positive because the number operators commute with the operators on the opposite subsystem.
On $\mathcal{H}_{\mathrm{eq}}$, the number-difference terms in Eq.~\eqref{eq:app-commutator-decomposition} cancel.
Since $\mathcal{X},\mathcal{Y} \geq 0$, we obtain $[P_{N},P_{N}^{\dagger}]\leq N\id$.

For the lower bound, fix a sector with exactly $t$ excitations on each side.
For a fixed set $R\subseteq\{0,\ldots,N-1\}$ of $t$ excited site indices in $A$, the operator $\mathcal{X}$ reduces to $\mathcal{X}_{R}:=\sum_{a\in R}q_{a}q_{a}^{\dagger}$ acting on the $t$-excitation subspace of $B$.
Set $(K_{R})_{bb'}:=\sum_{a\in R}\overline{(F_{N})_{ab}}(F_{N})_{ab'}$.
The rows of $F_{N}$ are orthonormal, so $K_{R}$ is a rank-$t$ orthogonal projection.
Using the spin commutation relation $\sigma_{j_{b}}^{-}\sigma_{j_{b'}}^{+} = \sigma_{j_{b'}}^{+}\sigma_{j_{b}}^{-}  + \delta_{bb'}(\id-2n_{j_{b}})$ and $\sum_{b}(K_{R})_{bb}=|R|=t$, we obtain
\begin{align}
	\mathcal{X}_{R} &= t\id + \sum_{b,b'}(K_{R}^{\mathsf{T}})_{bb'} \sigma_{j_{b}}^{+}\sigma_{j_{b'}}^{-} - 2 \sum_{b}(K_{R})_{bb} n_{j_{b}}.
	\label{eq:app-X-normal-order}
\end{align}
The transpose $K_{R}^{\mathsf{T}}$ is also an orthogonal projection, and the last term in Eq.~\eqref{eq:app-X-normal-order} is nonpositive.
Since $\mathcal{X}_{R}\geq0$ by its definition, Eq.~\eqref{eq:positive-matrix-spin-bound} gives $0\leq\mathcal{X}_{R}\leq t\id+\mathcal{N}_{B}=2t\id$ on the $t$-excitation subspace of $B$.
Consequently, $\mathcal{X}\leq2T$ on $\mathcal{H}_{\mathrm{eq}}$.

Similarly, on a $B$ configuration $S$ with $|S|=t$,
\begin{align}
	\mathcal{Y}_{S} = \sum_{b\in S} e_{b}^{\dagger}e_{b} = \sum_{a,a'}(L_{S})_{aa'}\sigma_{i_{a}}^{+}\sigma_{i_{a'}}^{-},
	\label{eq:app-Y-S}
\end{align}
where
\begin{align}
	(L_{S})_{aa'}:=\sum_{b\in S}(F_{N})_{ab}\overline{(F_{N})_{a'b}}.
	\label{eq:app-L-S}
\end{align}
Column orthogonality makes $L_{S}$ a rank-$t$ orthogonal projection.
Equation~\eqref{eq:positive-matrix-spin-bound} gives $0\leq\mathcal{Y}_{S}\leq\mathcal{N}_{A}=t\id$, and hence $\mathcal{Y}\leq T$ on $\mathcal{H}_{\mathrm{eq}}$.
Substituting $\mathcal{X}\leq2T$ and $\mathcal{Y} \leq T$ into Eq.~\eqref{eq:app-commutator-decomposition} gives
\begin{equation}
	N\id-6T\leq[P_{N},P_{N}^{\dagger}]\leq N\id
	\qquad\text{on }\mathcal{H}_{\mathrm{eq}}.
	\label{eq:app-pair-commutator-bound}
\end{equation}

\subsection{Differential Bounds along the Unitary Trial Flow}

Recall the trial state $\ket{\psi_{s}}=\ee^{s(P_{N}^{\dagger}-P_{N})}\ket{\mathbf{0}}$ from Eq.~\eqref{eq:trial-flow-main}.
The operator $P_{N}^{\dagger}-P_{N}$ is anti-Hermitian, so the trial state is normalized.
Moreover, $\mathcal{D}\ket{\mathbf{0}}=0$ and $[P_{N}^{\dagger}-P_{N},\mathcal{D}]=0$ imply that $\ket{\psi_{s}}\in\mathcal{H}_{\mathrm{eq}}$ for all $s$.
For $m(s):=\bra{\psi_{s}}T\ket{\psi_{s}}$, the definition of the trial state gives $\frac{\dd}{\dd s}\ket{\psi_{s}}=(P_{N}^{\dagger}-P_{N})\ket{\psi_{s}}$ and $\frac{\dd}{\dd s}\bra{\psi_{s}}=-\bra{\psi_{s}}(P_{N}^{\dagger}-P_{N})$.
Thus, differentiating $m(s)$ produces the commutator $[T,P_{N}^{\dagger}-P_{N}]$.
Since $P_{N}^{\dagger}$ increases the excitation number on $A$ by one and $P_{N}$ decreases it by one, we have $[T,P_{N}^{\dagger}]=P_{N}^{\dagger}$ and $[T,P_{N}]=-P_{N}$ on $\mathcal{H}_{\mathrm{eq}}$.
Consequently,
\begin{equation}
	m'(s)=\bra{\psi_{s}}(P_{N}^{\dagger}+P_{N})\ket{\psi_{s}}.
	\label{eq:app-flow-first-derivative}
\end{equation}
Differentiating once more and using $[P_{N}^{\dagger}+P_{N},P_{N}^{\dagger}-P_{N}]=2[P_{N},P_{N}^{\dagger}]$, we obtain
\begin{equation}
	m''(s)=2\bra{\psi_{s}}[P_{N},P_{N}^{\dagger}]\ket{\psi_{s}}.
	\label{eq:app-flow-second-derivative}
\end{equation}
Taking expectations of Eq.~\eqref{eq:app-pair-commutator-bound} in $\ket{\psi_{s}}$ gives
\begin{equation}
	2N-12m(s)\leq m''(s)\leq2N.
	\label{eq:app-flow-differential-bound}
\end{equation}
The initial conditions are $m(0)=m'(0)=0$.

We first derive the upper bound on $m(s)$.
Integrating the upper inequality in Eq.~\eqref{eq:app-flow-differential-bound} twice and using the initial conditions gives
\begin{equation}
	m(s)=\int_{0}^{s}(s-u)m''(u)\,\dd u\leq Ns^{2}. 
	\label{eq:app-flow-occupation-upper}
\end{equation}

For the lower bound on $m'(s)$, we combine the lower inequality in Eq.~\eqref{eq:app-flow-differential-bound} with the upper bound on $m(s)$ in Eq.~\eqref{eq:app-flow-occupation-upper}.
For $s\geq0$, this gives
\begin{equation}
	m''(s)\geq2N-12m(s)\geq2N-12Ns^{2}.
\end{equation}
Integrating this inequality from $0$ to $s$ and using $m'(0)=0$, we obtain
\begin{align}
	m'(s)
	&=\int_{0}^{s}m''(u)\,\dd u \notag \\
	&\geq\int_{0}^{s}(2N-12Nu^{2})\,\dd u \notag \\
	&=2Ns-4Ns^{3}.
	\label{eq:app-flow-derivative-lower}
\end{align}

\subsection{Completing the Energy-Defect Estimate}

Since $\ket{\psi_{s}} \in \mathcal{H}_{\mathrm{eq}}$ and $\mathcal{N}=2T$ on this subspace, $\bra{\psi_{s}}H_{N}(g)\ket{\psi_{s}}=2m(s)-gm'(s)$.
For $0\leq g \leq 1/64$, evaluating the variational energy at $s=g/2$ and using Eqs.~\eqref{eq:app-flow-occupation-upper} and \eqref{eq:app-flow-derivative-lower} gives
\begin{align}
	E_{N}(g)
	&\leq2m\br{\frac{g}{2}}-gm'\br{\frac{g}{2}} \notag \\
	&\leq\frac{1}{2}Ng^{2}-g\br{Ng-\frac{1}{2}Ng^{3}} \notag \\
	&=-\frac{1}{2}Ng^{2}+\frac{1}{2}Ng^{4}.
\end{align}
The reference lower bound is $E_{N}(g)\geq Ne_{\mathrm{ref}}(g)\geq- Ng^{2} / 2$.
Subtracting $Ne_{\mathrm{ref}}(g)$ from the variational upper bound therefore gives
\begin{equation}
	0\leq E_{N}(g)-Ne_{\mathrm{ref}}(g)\leq\frac{1}{2}Ng^{4}.
\end{equation}
This proves Proposition~\ref{prop:energy-defect} with constants independent of $N$.

\section{Excitation-Number Bounds and the Entropy Upper Bound}
\label{app:upper}

In this appendix, we provide the detailed excitation-number estimates used to establish the entropy upper bound in Theorem~\ref{thm:finite-size-main}.
Fix any normalized pure ground state $\ket{\Omega}$ of $H_{N}(g)$ for $0<g\leq1/64$, and let $\ket{\widetilde{\Omega}}:=J\ket{\Omega}$.
Recall $m_{A}:=\bra{\Omega}\mathcal{N}_{A}\ket{\Omega}$ and $m_{B}:=\bra{\Omega}\mathcal{N}_{B}\ket{\Omega}$ from Sec.~\ref{subsec:entropy-upper}.

By the exact compression relation in Eq.~\eqref{eq:exact-compression-expectation}, the reference energy expectation in $\ket{\widetilde{\Omega}}$ equals $E_{N}(g)$.
The first inequality in Eq.~\eqref{eq:reference-gap} therefore bounds the total reference-channel excitation weight by $4\delta_{N}(g)/3$.
Combining the reference gap in Eq.~\eqref{eq:reference-gap} with Proposition~\ref{prop:energy-defect}, we obtain
\begin{equation}
	\sum_{b}\bra{\widetilde{\Omega}}Q_{b}(r(g))\ket{\widetilde{\Omega}}
	\leq\frac{4}{3}\delta_{N}(g)\leq\frac{2}{3}Ng^{4}.
	\label{eq:app-reference-defect-g}
\end{equation}
Using $g=2r(g)/(1-r(g)^{2})$ from Eq.~\eqref{eq:g-r-relations}, we further obtain
\begin{equation}
	\sum_{b}\bra{\widetilde{\Omega}}Q_{b}(r(g))\ket{\widetilde{\Omega}}
	\leq\frac{32Nr(g)^{4}}{3(1-r(g)^{2})^{4}}
	\leq11Nr(g)^{4}.
	\label{eq:app-reference-defect}
\end{equation}
The last inequality follows from $r(g)\leq1/128$, which gives $32/[3(1-r(g)^{2})^{4}]\leq11$.

\subsection{Local Excitation Estimate}

We first establish a general operator inequality that relates a positive contraction to its expectation in a rank-one subspace.
Let $\mathcal{P}$ be a rank-one projection, let $\mathcal{Q}:=\id-\mathcal{P}$, and suppose that $0\leq X\leq\id$ and $\mathcal{P}X\mathcal{P}=p\mathcal{P}$.
Consider an arbitrary vector $\ket{u}+\ket{v}$, where $\ket{u}\in\Ran\mathcal{P}$ and $\ket{v}\in\Ran\mathcal{Q}$.
The assumptions on $X$ imply $\norm{X^{\frac{1}{2}}\ket{u}}^{2}=p\norm{\ket{u}}^{2}$ and $\norm{X^{\frac{1}{2}}\ket{v}}^{2}\leq\norm{\ket{v}}^{2}$.
Applying the triangle inequality to $X^{\frac{1}{2}}(\ket{u}+\ket{v})$ and using $(a+b)^{2}\leq2a^{2}+2b^{2}$ gives
\begin{equation}
	\bra{u+v}X\ket{u+v}
	\leq2p\norm{\ket{u}}^{2}+2\norm{\ket{v}}^{2}.
	\label{eq:app-local-projection-estimate}
\end{equation}
Since this holds for every vector, we obtain the operator inequality $X\leq2p\mathcal{P}+2\mathcal{Q}\leq2p\id+2\mathcal{Q}$.

We now apply this estimate to the $b$th reference channel.
Take $X=n_{j_{b}}$ and $\mathcal{P}=\ket{\varphi_{r(g)}}\bra{\varphi_{r(g)}}_{b}$, so that $\mathcal{Q}=Q_{b}(r(g))$.
The reference ground state in Eq.~\eqref{eq:reference-local-ground} satisfies $\mathcal{P}n_{j_{b}}\mathcal{P}=p\mathcal{P}$ with $p=r(g)^{2}/(1+r(g)^{2})$.
The general operator inequality therefore gives
\begin{equation}
	n_{j_{b}}\leq\frac{2r(g)^{2}}{1+r(g)^{2}}\id+2Q_{b}(r(g)).
	\label{eq:app-local-number-bound}
\end{equation}
This is precisely Eq.~\eqref{eq:local-number-bound} in the main text.

Taking expectations of Eq.~\eqref{eq:app-local-number-bound} in $\ket{\widetilde{\Omega}}$ and summing over $b$, the left-hand side gives $m_{B}$, since the embedding $J$ acts only on subsystem $A$.
Using Eq.~\eqref{eq:app-reference-defect} to bound the sum of the reference-channel excitation terms, we obtain
\begin{equation}
	m_{B}\leq\frac{2Nr(g)^{2}}{1+r(g)^{2}}+22Nr(g)^{4}
	\leq3Nr(g)^{2}.
	\label{eq:app-mB}
\end{equation}
The last inequality follows from $1/(1+r(g)^{2})\leq1$ and $22r(g)^{2}\leq1$, where the latter holds because $r(g)\leq1/128$.

\subsection{Exact Sum-of-Squares Identity and Number Imbalance}

To control the excitation number on $A$, we use the operators $\Lambda_{b}$ and $\Gamma_{a}$ defined in Eqs.~\eqref{eq:SOS-Lambda} and \eqref{eq:SOS-Gamma}.
We evaluate the sums of their squared norms by expanding them in powers of $r(g)$.
The terms independent of $r(g)$ give $\mathcal{N}_{B}$ for $\sum_{b}\Lambda_{b}^{\dagger}\Lambda_{b}$ and $\mathcal{N}_{A}$ for $\sum_{a}\Gamma_{a}^{\dagger}\Gamma_{a}$.
Since spin operators on different subsystems commute, the terms linear in $r(g)$ combine into $-r(g)(P_{N}^{\dagger}+P_{N})$ in both expansions, using Eq.~\eqref{eq:P-def}.

For the quadratic term in $\sum_{b}\Lambda_{b}^{\dagger}\Lambda_{b}$, the row orthogonality of $F_{N}$ gives $\sum_{b}\overline{(F_{N})_{ab}}(F_{N})_{a'b}=\delta_{aa'}$.
Together with $\sigma_{i_{a}}^{-}\sigma_{i_{a}}^{+}=\id-n_{i_{a}}$, this yields
\begin{align}
	\sum_{a,a',b}\overline{(F_{N})_{ab}}(F_{N})_{a'b}
	\sigma_{i_{a}}^{-}\sigma_{i_{a'}}^{+} = N\id-\mathcal{N}_{A}.
	\label{eq:app-Lambda-quadratic}
\end{align}
Similarly, the quadratic term in $\sum_{a}\Gamma_{a}^{\dagger}\Gamma_{a}$ is evaluated using the column orthogonality $\sum_{a}\overline{(F_{N})_{ab}}(F_{N})_{ab'}=\delta_{bb'}$.
Since $\sigma_{j_{b}}^{-}\sigma_{j_{b}}^{+}=\id-n_{j_{b}}$, we obtain
\begin{align}
	\sum_{a,b,b'}\overline{(F_{N})_{ab}}(F_{N})_{ab'}
	\sigma_{j_{b}}^{-}\sigma_{j_{b'}}^{+} = N\id-\mathcal{N}_{B}.
	\label{eq:app-Gamma-quadratic}
\end{align}
Combining the contributions at each order in $r(g)$, we obtain
\begin{align}
	\sum_{b}\Lambda_{b}^{\dagger}\Lambda_{b}
	&=\mathcal{N}_{B}-r(g)(P_{N}^{\dagger}+P_{N}) + r(g)^{2}(N\id-\mathcal{N}_{A}),
	\label{eq:app-Lambda-squares}\\
	\sum_{a}\Gamma_{a}^{\dagger}\Gamma_{a}
	&=\mathcal{N}_{A}-r(g)(P_{N}^{\dagger}+P_{N}) + r(g)^{2}(N\id-\mathcal{N}_{B}).
	\label{eq:app-Gamma-squares}
\end{align}

Adding Eqs.~\eqref{eq:app-Lambda-squares} and \eqref{eq:app-Gamma-squares}, we obtain
\begin{align}
	&\sum_{b}\Lambda_{b}^{\dagger}\Lambda_{b}
	+\sum_{a}\Gamma_{a}^{\dagger}\Gamma_{a} \notag \\
	&\quad=(1-r(g)^{2})\mathcal{N}
	-2r(g)(P_{N}^{\dagger}+P_{N})  +2Nr(g)^{2}\id \notag \\
	&\quad=(1-r(g)^{2})
	\br{H_{N}(g)-Ne_{\mathrm{ref}}(g)\id}.
	\label{eq:app-SOS-sum}
\end{align}
The last equality follows from $g=2r(g)/(1-r(g)^{2})$ and $e_{\mathrm{ref}}(g)=-2r(g)^{2}/(1-r(g)^{2})$ in Eq.~\eqref{eq:g-r-relations}. 
On the other hand, subtracting Eq.~\eqref{eq:app-Lambda-squares} from Eq.~\eqref{eq:app-Gamma-squares} gives
\begin{equation}
	\sum_{a}\Gamma_{a}^{\dagger}\Gamma_{a}
	-\sum_{b}\Lambda_{b}^{\dagger}\Lambda_{b}
	=(1+r(g)^{2})(\mathcal{N}_{A}-\mathcal{N}_{B}).
	\label{eq:app-SOS-difference}
\end{equation}

Taking the expectation of Eq.~\eqref{eq:app-SOS-difference} in $\ket{\Omega}$, we express $(1+r(g)^{2})(m_{A}-m_{B})$ as the difference of the two nonnegative quantities $\sum_{a}\norm{\Gamma_{a}\ket{\Omega}}^{2}$ and $\sum_{b}\norm{\Lambda_{b}\ket{\Omega}}^{2}$.
The absolute value of this difference is bounded by their sum.
Taking the expectation of Eq.~\eqref{eq:app-SOS-sum} in $\ket{\Omega}$ also shows that their sum equals $(1-r(g)^{2})\delta_{N}(g)$.
Combining these results, we obtain
\begin{equation}
	(1+r(g)^{2})|m_{A}-m_{B}|
	\leq(1-r(g)^{2})\delta_{N}(g).
	\label{eq:app-number-imbalance}
\end{equation}
Dividing by $1+r(g)^{2}$ proves Eq.~\eqref{eq:imbalance-main}.

We now combine the imbalance estimate with the bound on $m_{B}$ in Eq.~\eqref{eq:app-mB} to obtain an upper bound on $m_{A}$.
By Proposition~\ref{prop:energy-defect} and the relation between $g$ and $r(g)$ in Eq.~\eqref{eq:g-r-relations},
\begin{equation}
	\delta_{N}(g)
	\leq\frac{1}{2}Ng^{4}
	=\frac{8Nr(g)^{4}}{(1-r(g)^{2})^{4}}
	\leq9Nr(g)^{4},
	\label{eq:app-energy-defect-r-bound}
\end{equation}
where the last inequality again uses $r(g)\leq1/128$.
Since $(1-r(g)^{2})/(1+r(g)^{2})\leq1$, Eq.~\eqref{eq:app-number-imbalance} implies $|m_{A}-m_{B}|\leq\delta_{N}(g)$.
Thus, $m_{A}\leq m_{B}+\delta_{N}(g)$, and Eqs.~\eqref{eq:app-mB} and \eqref{eq:app-energy-defect-r-bound} give
\begin{equation}
	m_{A}\leq3Nr(g)^{2}+9Nr(g)^{4}
	\leq4Nr(g)^{2}.
	\label{eq:app-mA}
\end{equation}
The last inequality follows from $9r(g)^{2}\leq1$, which holds under the assumed coupling condition.
Hence, the excitation densities on both subsystems are bounded by quantities of order $r(g)^{2}$.

\subsection{From Excitation Density to Entropy}

We finally convert the excitation-number bound into an upper bound on the bipartite entanglement entropy.
For each $i_{a}\in A$, let $p_{a}:=\bra{\Omega}n_{i_{a}}\ket{\Omega}$ and let $\rho_{i_{a}}$ denote the corresponding one-qubit reduced state.
Dephasing $\rho_{i_{a}}$ in the computational basis gives a diagonal state with eigenvalues $1-p_{a}$ and $p_{a}$.
Since dephasing is an average of unitary conjugations, concavity and unitary invariance of the von Neumann entropy imply $S(\rho_{i_{a}})\leq h_{2}(p_{a})$.

The subadditivity of the von Neumann entropy bounds the entropy of subsystem $A$ by the sum of its one-qubit entropies.
Applying the preceding one-qubit estimate and then using the concavity of $h_{2}$, together with $\sum_{a}p_{a}=m_{A}$, gives
\begin{align}
	S_{A}(\ket{\Omega}) \leq \sum_{a}S(\rho_{i_{a}}) \leq \sum_{a}h_{2}(p_{a}) \leq N h_{2} \left(\frac{m_{A}}{N}\right).
	\label{eq:app-entropy-concavity}
\end{align}
By Eq.~\eqref{eq:app-mA}, the mean excitation density satisfies $m_{A}/N\leq4r(g)^{2}$.
Moreover, $r(g)\leq1/128$ ensures that $4r(g)^{2}<1/2$.
Since the binary entropy function $h_{2}$ is increasing on $[0,1/2]$, Eq.~\eqref{eq:app-entropy-concavity} therefore yields
\begin{equation}
	S_{A}(\ket{\Omega})\leq Nh_{2}(4r(g)^{2}).
	\label{eq:app-entropy-upper-bound}
\end{equation}
This proves the upper bound in Theorem~\ref{thm:finite-size-main}.

\section{Entanglement Inequalities and One-Pair Block Extraction}
\label{app:entanglement-tools}

In this appendix, we review the properties of squashed entanglement and entropy continuity and provide the detailed proof of Proposition~\ref{prop:block-extraction}.

\subsection{Squashed Entanglement and Coherent Information}
 
Let $\rho_{AB}$ be a finite-dimensional bipartite state.
Strong subadditivity of the von Neumann entropy implies that $I(A:B|E)_{\rho_{ABE}}\geq0$ for every extension $\rho_{ABE}$ of $\rho_{AB}$~\cite{LiebRuskai1973}.
By the definition of squashed entanglement in Eq.~\eqref{eq:squashed-definition}, this gives $E_{\sq}(A:B)_{\rho_{AB}} \geq 0$.
For a pure bipartite state $\ket{\psi}_{AB}$, every extension has the product form $\ket{\psi}\bra{\psi}_{AB}\otimes\rho_{E}$.
Its conditional mutual information equals $2S(\rho_{A})$, where $\rho_{A}:=\Tr_{B}(\ket{\psi}\bra{\psi})$.
Consequently,
\begin{equation}
	E_{\sq}(A:B)_{\psi}=S(\rho_{A}).
	\label{eq:sq-pure-entropy}
\end{equation}

We use the strong superadditivity of squashed entanglement~\cite{ChristandlWinter}.
For a state $\rho$ on $A_{1}\cdots A_{M}B_{1}\cdots B_{M}$, this property states that
\begin{equation}
	E_{\sq}(A_{1}\cdots A_{M}:B_{1}\cdots B_{M})_{\rho}
	\geq\sum_{\ell=1}^{M}E_{\sq}(A_{\ell}:B_{\ell})_{\rho_{A_{\ell}B_{\ell}}}.
	\label{eq:sq-strong-superadditivity}
\end{equation}
Here $\rho_{A_{\ell}B_{\ell}}$ is the reduced state on the corresponding pair of subsystems.

We also use the strong monotonicity of squashed entanglement on average under local operations and classical communication (LOCC)~\cite{ChristandlWinter}.
For an LOCC protocol with outcomes $x$ occurring with probabilities $p_{x}$ and corresponding normalized output states $\rho^{(x)}_{A'B'}$, this property gives
\begin{equation}
	E_{\sq}(A:B)_{\rho_{AB}} \geq \sum_{x}p_{x}E_{\sq}(A':B')_{\rho^{(x)}}.
	\label{eq:app-sq-strong-monotonicity}
\end{equation}
Since squashed entanglement is nonnegative, the contribution from any postselected outcome, multiplied by its success probability, cannot exceed the initial entanglement.
Discarding local subsystems also cannot increase squashed entanglement.

Finally, we use the coherent-information lower bound~\cite{ChristandlWinter,DevetakWinter}
\begin{equation}
	E_{\sq}(A:B)_{\rho_{AB}} \geq \max\{0,S(\rho_{B})-S(\rho_{AB})\}.
	\label{eq:sq-coherent-information}
\end{equation}
This bound follows from the one-way hashing inequality and the fact that squashed entanglement upper-bounds distillable entanglement.

\subsection{Trace Distance and Entropy Continuity}

For density operators $\rho$ and $\sigma$, define their trace distance by $d_{\mathrm{tr}}(\rho,\sigma):=\norm{\rho-\sigma}_{1}/2$.
The Fuchs--van de Graaf inequality gives the following bound when one of the states is pure~\cite{FuchsVandeGraaf}:
\begin{equation}
	d_{\mathrm{tr}}\br{\rho,\ket{\phi}\bra{\phi}} \leq \sqrt{1-\bra{\phi}\rho\ket{\phi}}.
	\label{eq:pure-trace-bound}
\end{equation}  
For any bipartite density operators $\rho_{AB}$ and $\sigma_{AB}$, the contractivity of trace distance under partial trace gives
\begin{equation}
	d_{\mathrm{tr}}(\Tr_{B}\rho_{AB},\Tr_{B}\sigma_{AB}) \leq d_{\mathrm{tr}}(\rho_{AB},\sigma_{AB}).
	\label{eq:app-trace-contractivity}
\end{equation}
Moreover, for any density operators $\rho$ and $\sigma$ and any measurement operator $0\leq M\leq\id$, the difference between the corresponding outcome probabilities satisfies
\begin{equation}
	|\Tr(M\rho)-\Tr(M\sigma)| \leq d_{\mathrm{tr}}(\rho,\sigma).
	\label{eq:app-measurement-probability}
\end{equation}

For two states $\rho$ and $\sigma$ supported on a common $d$-dimensional Hilbert space, the Fannes--Audenaert inequality implies~\cite{Audenaert}
\begin{align}
	|S(\rho)-S(\sigma)| &\leq d_{\mathrm{tr}}(\rho,\sigma)\log d + h_{2}(d_{\mathrm{tr}}(\rho,\sigma)).
	\label{eq:entropy-continuity-tool}
\end{align}
Here the sharper factor $\log(d-1)$ has been replaced by $\log d$, which is sufficient for our estimates.

\subsection{Proof of the One-Pair Extraction Proposition} 

We now prove Proposition~\ref{prop:block-extraction}.
Let $G\subseteq\{0,\ldots,N-1\}$ be a set of reference-channel indices with $|G|=L$, and let $\sigma_{G}$ be a density operator on $\mathcal{F}_{A_{G}}\otimes\mathcal{H}_{B_{G}}$ satisfying the assumptions of the proposition.
For convenience, relabel the channels in $G$ by $b=1,\ldots,L$.
Recall from Eqs.~\eqref{eq:reference-local-ground} and \eqref{eq:block-reference-and-probability} that $\ket{\Phi_{r,G}}=\bigotimes_{b\in G}\ket{\varphi_{r}}_{b}$, where $\ket{\varphi_{r}}_{b}=(\ket{0,0}_{b}+r\ket{1,1}_{b})/\sqrt{1+r^{2}}$.
The corresponding one-pair probability is $p_{L}(r)=Lr^{2}/(1+r^{2})^{L}$.
The total reference-channel excitation weight is $\epsilon_{G}:=\sum_{b\in G}\Tr(\sigma_{G}Q_{b}(r))$, which satisfies $\epsilon_{G}\leq p_{L}(r)^{2}/100$ by Eq.~\eqref{eq:block-good-condition}.

Since the projectors $Q_{b}(r)$ commute and act on distinct channels, their joint spectrum gives
\begin{align}
	\id-\ket{\Phi_{r,G}}\bra{\Phi_{r,G}} = \id-\prod_{b=1}^{L}(\id-Q_{b}(r)) \leq \sum_{b=1}^{L}Q_{b}(r).
	\label{eq:app-block-reference-projector}
\end{align}
Taking the expectation in $\sigma_{G}$ therefore gives $\bra{\Phi_{r,G}}\sigma_{G}\ket{\Phi_{r,G}}\geq1-\epsilon_{G}$.
Using Eq.~\eqref{eq:pure-trace-bound} and the hypothesis in Eq.~\eqref{eq:block-good-condition}, we obtain
\begin{align}
	d_{\mathrm{tr}}\br{\sigma_{G},\ket{\Phi_{r,G}}\bra{\Phi_{r,G}}} \leq \sqrt{\epsilon_{G}} \leq \frac{p_{L}(r)}{10}.
	\label{eq:block-trace-comparison}
\end{align}

We now consider local projections onto exactly one excitation on each side of the selected channels.
For $b=1,\ldots,L$, let $\ket{b}_{A}$ denote the state with one boson in mode $d_{b}$ and no bosons in the other selected modes.
Likewise, let $\ket{b}_{B}$ denote the state with only spin $j_{b}$ excited among the selected $B$ spins.
Define
\begin{equation}
	R_{A}:=\sum_{b=1}^{L}\ket{b}_{A}\bra{b}_{A}, \quad R_{B}:=\sum_{b=1}^{L}\ket{b}_{B}\bra{b}_{B}.
	\label{eq:block-local-projectors}
\end{equation}
We also set $R:=R_{A}\otimes R_{B}$.
By Eqs.~\eqref{eq:reference-local-ground} and \eqref{eq:block-reference-and-probability}, the component of $\ket{\Phi_{r,G}}$ containing exactly one excitation pair is obtained by selecting the $r\ket{1,1}_{b}$ term from one channel and the vacuum term from all remaining channels.
This gives an amplitude $r/(1+r^{2})^{L/2}$ for each of the $L$ channels.
Consequently,
\begin{equation}
	R\ket{\Phi_{r,G}} = \sqrt{p_{L}(r)}\ket{\Phi_{L}},
	\label{eq:block-ideal-postselection}
\end{equation}
where
\begin{equation}
	\ket{\Phi_{L}} := \frac{1}{\sqrt{L}}\sum_{b=1}^{L}\ket{b}_{A}\ket{b}_{B}.
	\label{eq:block-maximally-entangled-state}
\end{equation}
Thus, applying the projection $R$ to the reference state $\ket{\Phi_{r,G}}$ succeeds with probability $p_{L}(r)=Lr^{2}/(1+r^{2})^{L}$ and produces the maximally entangled state $\ket{\Phi_{L}}$ of Schmidt rank $L$.

For the actual state $\sigma_{G}$, let $p':=\Tr(R\sigma_{G})$ be the success probability of the same projection.
The measurement probability bound in Eq.~\eqref{eq:app-measurement-probability}, together with Eq.~\eqref{eq:block-trace-comparison}, gives $|p'-p_{L}(r)|\leq p_{L}(r)/10$.
Therefore,
\begin{equation}
	p'\geq p_{L}(r)-\frac{p_{L}(r)}{10} = \frac{9}{10}p_{L}(r) > 0.
	\label{eq:block-success-bound}
\end{equation}
The normalized postselected state is
\begin{equation}
	\omega:=\frac{R\sigma_{G}R}{p'}.
	\label{eq:block-actual-postselection}
\end{equation}

We next compare $\omega$ with $\ket{\Phi_{L}}$.
Equation~\eqref{eq:block-ideal-postselection} implies $R\ket{\Phi_{r,G}}\bra{\Phi_{r,G}}R=p_{L}(r)\ket{\Phi_{L}}\bra{\Phi_{L}}$.
Since projection cannot increase the trace norm, Eq.~\eqref{eq:block-trace-comparison} gives
\begin{equation}
	\norm{R\br{\sigma_{G}-\ket{\Phi_{r,G}}\bra{\Phi_{r,G}}}R}_{1} \leq \frac{p_{L}(r)}{5}.
	\label{eq:app-block-projected-distance}
\end{equation}  
Substituting Eqs.~\eqref{eq:block-actual-postselection} and \eqref{eq:block-ideal-postselection} into Eq.~\eqref{eq:app-block-projected-distance}, we obtain
\begin{equation}
	\norm{p'\omega-p_{L}(r)\ket{\Phi_{L}}\bra{\Phi_{L}}}_{1} \leq \frac{p_{L}(r)}{5}.
	\label{eq:app-block-unnormalized-comparison}
\end{equation}
To bound the trace distance between $\omega$ and $\ket{\Phi_{L}}$, we must also account for the difference between the success probabilities $p'$ and $p_{L}(r)$.
Adding and subtracting $p_{L}(r)\ket{\Phi_{L}}\bra{\Phi_{L}}$ in $p'(\omega-\ket{\Phi_{L}}\bra{\Phi_{L}})$, the triangle inequality bounds its trace norm by $p_{L}(r)/5+|p'-p_{L}(r)|$.
Since $|p'-p_{L}(r)|\leq p_{L}(r)/10$ and $p'\geq9p_{L}(r)/10$ by Eq.~\eqref{eq:block-success-bound}, we obtain
\begin{equation}
	d_{\mathrm{tr}}\br{\omega,\ket{\Phi_{L}}\bra{\Phi_{L}}} \leq \frac{3p_{L}(r)}{20p'} < \frac{2}{9}.
	\label{eq:block-output-distance}
\end{equation}

We now apply entropy continuity to obtain a lower bound on the entanglement of $\omega$.
Since $\omega=R\sigma_{G}R/p'$ and $R=R_{A}\otimes R_{B}$, the postselected state is supported on the one-excitation subspaces of $A_{G}$ and $B_{G}$, each of dimension $L$.
Thus, the range of $R$ has dimension $L^{2}$.
By the Schmidt decomposition in Eq.~\eqref{eq:block-maximally-entangled-state}, tracing out $A_{G}$ from $\ket{\Phi_{L}}\bra{\Phi_{L}}$ gives the reduced state $R_{B}/L$ on $B_{G}$, which has entropy $\log L$.
Since $\ket{\Phi_{L}}$ is pure, its joint von Neumann entropy is zero.

Let $\omega_{B_{G}}:=\Tr_{A_{G}}\omega$.
Since the reduced state of $\ket{\Phi_{L}}$ on $B_{G}$ is $R_{B}/L$, the trace-distance contractivity in Eq.~\eqref{eq:app-trace-contractivity} and Eq.~\eqref{eq:block-output-distance} give $d_{\mathrm{tr}}(\omega_{B_{G}},R_{B}/L)\leq2/9$.
Both reduced states are supported on the $L$-dimensional range of $R_{B}$, and $S(R_{B}/L)=\log L$.
Applying the entropy continuity inequality in Eq.~\eqref{eq:entropy-continuity-tool} with $d=L$, we obtain
\begin{equation}
	S(\omega_{B_{G}}) \geq \frac{7}{9}\log L - h_{2}\br{\frac{2}{9}}.
	\label{eq:app-block-marginal-entropy}
\end{equation}
For the joint state, Eq.~\eqref{eq:block-output-distance} directly bounds the trace distance between $\omega$ and $\ket{\Phi_{L}}\bra{\Phi_{L}}$ by $2/9$.
Both $\omega$ and $\ket{\Phi_{L}}\bra{\Phi_{L}}$ are supported on the range of $R$, which has dimension $L^{2}$, and $S(\ket{\Phi_{L}}\bra{\Phi_{L}})=0$.
Applying Eq.~\eqref{eq:entropy-continuity-tool} with $d=L^{2}$ therefore gives
\begin{equation}
	S(\omega) \leq \frac{4}{9}\log L + h_{2}\br{\frac{2}{9}}.
	\label{eq:app-block-joint-entropy}
\end{equation}
In both applications, we used the fact that $x\log d+h_{2}(x)$ is increasing on $[0,2/9]$ for every $d \geq 1$.

Combining Eqs.~\eqref{eq:app-block-marginal-entropy} and \eqref{eq:app-block-joint-entropy}, we obtain
\begin{align}
	S(\omega_{B_{G}}) - S(\omega) \geq \frac{1}{3}\log L - 2h_{2}\br{\frac{2}{9}} \geq \frac{1}{5}\log L.
	\label{eq:block-coherent-information}
\end{align}
The last inequality follows from $L \geq 4096$, which ensures $15h_{2}(2/9)<\log L$.
Applying the coherent-information lower bound in Eq.~\eqref{eq:sq-coherent-information}, we therefore obtain
\begin{equation}
	E_{\sq}(A_{G}:B_{G})_{\omega} \geq \frac{1}{5}\log L.
	\label{eq:app-block-postselected-entanglement}
\end{equation}

Finally, the local measurements $\{R_{A},\id-R_{A}\}$ and $\{R_{B},\id-R_{B}\}$ constitute an LOCC protocol whose joint successful outcome occurs with probability $p'$.
Applying Eq.~\eqref{eq:app-sq-strong-monotonicity} to this LOCC protocol and dropping the nonnegative contributions from all other outcomes gives
\begin{align}
	E_{\sq}(A_{G}:B_{G})_{\sigma_{G}}
	&\geq p'E_{\sq}(A_{G}:B_{G})_{\omega} \notag \\
	&\geq\frac{9}{50}p_{L}(r)\log L.
	\label{eq:app-block-extraction-final}
\end{align}
The last inequality follows from Eqs.~\eqref{eq:block-success-bound} and \eqref{eq:app-block-postselected-entanglement}.
This proves Proposition~\ref{prop:block-extraction}.

\section{Global Entanglement Lower Bound}
\label{app:global-lower}

In this appendix, we complete the proof of the entropy lower bound in Theorem~\ref{thm:finite-size-main}.
Assume its hypotheses, namely $0<\gamma\leq1/64$, $1/2\leq\kappa\leq1$, and $N\geq4096$.
Then $g_{N,\kappa} = \gamma N^{\frac{1}{2}-\kappa} \leq \gamma \leq 1/64$ and $0 < r_{N,\kappa} < g_{N,\kappa}/2 \leq 1/128$.
Recall the block size defined in Eq.~\eqref{eq:L-def},
\begin{equation}
	L_{N,\kappa} = \min\brrr{N, \left\lfloor\frac{1}{2r_{N,\kappa}^{2}}\right\rfloor}.
	\label{eq:app-global-L}
\end{equation}

Let $\ket{\Omega_{N,\kappa}}$ be any normalized pure ground state of $H_{N,\kappa}$, and let $\ket{\widetilde{\Omega}_{N,\kappa}}:=J\ket{\Omega_{N,\kappa}}$ be its embedded state.
Since $J$ is an isometry acting only on subsystem $A$, the embedded state is pure and has the same Schmidt coefficients across the bipartition.
For a set $G\subseteq\{0,\ldots,N-1\}$ of reference-channel indices, let $\rho_{G}$ denote the reduced density operator of $\ket{\widetilde{\Omega}_{N,\kappa}}$ on $\mathcal{F}_{A_{G}}\otimes\mathcal{H}_{B_{G}}$, as defined in Sec.~\ref{subsec:entropy-lower}.
By Eq.~\eqref{eq:sq-pure-entropy}, we have
\begin{equation}
	S_{A}(\ket{\Omega_{N,\kappa}})
	=E_{\sq}(A:B)_{\widetilde{\Omega}_{N,\kappa}}.
	\label{eq:app-global-pure-sq}
\end{equation}
Here, $A$ denotes the bosonic subsystem with Hilbert space $\mathcal{F}_{A}$, while $B$ denotes the original spin subsystem with Hilbert space $\mathcal{H}_{B}$.

Applying Eq.~\eqref{eq:app-reference-defect} with $g=g_{N,\kappa}$, we control the total reference-channel excitation weight by $11Nr_{N,\kappa}^{4}$.
We treat separately the cases $L_{N,\kappa}<N$ and $L_{N,\kappa}=N$.

\subsection{Many Blocks with Order-One Pairing Weight}

Suppose first that $L_{N,\kappa} < N$, so that $L_{N,\kappa} = \lfloor(2r_{N,\kappa}^{2})^{-1}\rfloor$.
Define $M:=\lfloor N/L_{N,\kappa}\rfloor$ and choose $M$ disjoint sets of reference-channel indices $G_{1},\ldots,G_{M}$, each containing exactly $L_{N,\kappa}$ channels.
The remaining $N-ML_{N,\kappa}$ channels are not included in these blocks.
For each block $G_{\ell}$, define its total reference-channel excitation weight by
\begin{equation}
	\epsilon_{G_{\ell}}
	:=\sum_{b\in G_{\ell}}
	\Tr\br{\rho_{G_{\ell}}Q_{b}(r_{N,\kappa})}.
	\label{eq:app-block-defect}
\end{equation}
Because the blocks are disjoint, each reference-channel projector appears in at most one of these sums.
Moreover, each $Q_{b}(r_{N,\kappa})$ is positive semidefinite, so omitting the remaining channels cannot increase the total excitation weight.
Equation~\eqref{eq:app-reference-defect} therefore gives
\begin{equation}
	\sum_{\ell=1}^{M}\epsilon_{G_{\ell}}
	\leq11Nr_{N,\kappa}^{4}.
	\label{eq:app-sum-block-defects}
\end{equation}

We next show that, for each block $G_{\ell}$ of size $L_{N,\kappa}$, applying the projection $R$ to the reference state $\ket{\Phi_{r_{N,\kappa},G_{\ell}}}$ succeeds with a probability bounded below by a positive constant independent of $N$.
Since $L_{N,\kappa}=\lfloor(2r_{N,\kappa}^{2})^{-1}\rfloor$, the definition of the floor function gives $L_{N,\kappa}\leq(2r_{N,\kappa}^{2})^{-1}<L_{N,\kappa}+1$.
Multiplying these inequalities by $r_{N,\kappa}^{2}>0$, we obtain
\begin{equation}
	\frac{1}{2}-r_{N,\kappa}^{2}
	<L_{N,\kappa}r_{N,\kappa}^{2}
	\leq\frac{1}{2}.
	\label{eq:app-Lr2-window}
\end{equation}
Thus, $L_{N,\kappa}r_{N,\kappa}^{2}$ remains close to $1/2$, independently of $N$.
Recall from Eq.~\eqref{eq:block-reference-and-probability} that applying the projection $R$ to the reference state $\ket{\Phi_{r,G}}$, with $|G|=L$, succeeds with probability $p_{L}(r)=Lr^{2}/(1+r^{2})^{L}$.
Using $(1+r_{N,\kappa}^{2})^{L_{N,\kappa}}\leq\ee^{L_{N,\kappa}r_{N,\kappa}^{2}}\leq\ee^{1/2}$ together with Eq.~\eqref{eq:app-Lr2-window}, we obtain
\begin{align}
	p_{L_{N,\kappa}}(r_{N,\kappa})
	&\geq\br{\frac{1}{2}-r_{N,\kappa}^{2}}\ee^{-1/2}  > 0.30.
	\label{eq:app-dense-p}
\end{align}
The last inequality follows from $r_{N,\kappa}^{2} \leq 1/16384$.
This bound also implies $L_{N,\kappa} \geq 8192$, satisfying the condition $L \geq 4096$ in Proposition~\ref{prop:block-extraction}.

We now show that a sufficiently large fraction of these blocks satisfies the reference-channel excitation-weight condition in Eq.~\eqref{eq:block-good-condition} of Proposition~\ref{prop:block-extraction}.
Call a block $G_{\ell}$ good if
\begin{equation}
	\epsilon_{G_{\ell}}
	\leq\frac{p_{L_{N,\kappa}}(r_{N,\kappa})^{2}}{100},
	\label{eq:app-good-block-condition}
\end{equation}
and bad otherwise.  
Let $M_{\mathrm{bad}}$ denote the number of bad blocks.
By definition, every bad block contributes more than $p_{L_{N,\kappa}}(r_{N,\kappa})^{2}/100$ to the total sum in Eq.~\eqref{eq:app-sum-block-defects}.
Since that sum is at most $11Nr_{N,\kappa}^{4}$, the number of bad blocks must satisfy
\begin{align}
	M_{\mathrm{bad}}
	<\frac{1100Nr_{N,\kappa}^{4}}
	{p_{L_{N,\kappa}}(r_{N,\kappa})^{2}}  < 0.746Nr_{N,\kappa}^{2}.
	\label{eq:app-bad-count}
\end{align}
Here the second inequality follows from Eq.~\eqref{eq:app-dense-p} and $r_{N,\kappa}^{2}\leq1/16384$, which imply $1100r_{N,\kappa}^{2}/p_{L_{N,\kappa}}(r_{N,\kappa})^{2}<1100/(0.30^{2}\cdot16384)<0.746$.

To compare the number of bad blocks with the total number $M$, recall that $M=\lfloor N/L_{N,\kappa}\rfloor$.
Since $N/L_{N,\kappa}>1$, the elementary inequality $\lfloor x\rfloor\geq x/2$ for $x\geq1$ gives
\begin{align}
	M&\geq\frac{N}{2L_{N,\kappa}} \geq Nr_{N,\kappa}^{2},
	\label{eq:app-total-count}
\end{align}
where the last inequality uses $L_{N,\kappa}\leq(2r_{N,\kappa}^{2})^{-1}$.
Combining this with Eq.~\eqref{eq:app-bad-count}, the number of good blocks is bounded below by
\begin{align}
	M-M_{\mathrm{bad}}  >M-0.746Nr_{N,\kappa}^{2}  \geq0.254M>\frac{M}{4}.
	\label{eq:app-good-block-count}
\end{align}
Thus, at least one quarter of the selected blocks are good, independently of the system size.

For each good block $G_{\ell}$, the density operator $\rho_{G_{\ell}}$ satisfies Eq.~\eqref{eq:app-good-block-condition}, and its block size satisfies $L_{N,\kappa}\geq4096$.
Proposition~\ref{prop:block-extraction} therefore applies and gives a lower bound on its squashed entanglement.
Using the uniform probability bound in Eq.~\eqref{eq:app-dense-p}, we obtain
\begin{align}
	E_{\sq}(A_{G_{\ell}}:B_{G_{\ell}})_{\rho_{G_{\ell}}}
	&\geq\frac{9}{50}p_{L_{N,\kappa}}(r_{N,\kappa})
	\log L_{N,\kappa} \notag \\
	&>\frac{1}{20}\log L_{N,\kappa}.
	\label{eq:app-good-block-entanglement}
\end{align}
The last inequality holds because $(9/50)\cdot0.30=0.054>1/20$.
Hence, every good block contributes at least $(1/20)\log L_{N,\kappa}$ to the squashed-entanglement lower bound.

Since the embedding $J$ and the bosonic mode rotation act only on subsystem $A$, they preserve the entanglement across $A|B$.
Discarding unused reference channels cannot increase squashed entanglement.
Thus, Eq.~\eqref{eq:app-global-pure-sq} and the strong superadditivity of squashed entanglement in Eq.~\eqref{eq:sq-strong-superadditivity} give
\begin{equation}
	S_{A}(\ket{\Omega_{N,\kappa}})
	\geq\sum_{\ell=1}^{M}
	E_{\sq}(A_{G_{\ell}}:B_{G_{\ell}})_{\rho_{G_{\ell}}}.
	\label{eq:app-global-block-sum}
\end{equation}
By Eqs.~\eqref{eq:app-good-block-count} and \eqref{eq:app-good-block-entanglement}, at least $M/4$ blocks contribute at least $(1/20)\log L_{N,\kappa}$ each, while the remaining contributions are nonnegative.
Using $M\geq Nr_{N,\kappa}^{2}$ from Eq.~\eqref{eq:app-total-count}, we obtain
\begin{equation}
	S_{A}(\ket{\Omega_{N,\kappa}})
	\geq\frac{M}{80}\log L_{N,\kappa}
	\geq\frac{Nr_{N,\kappa}^{2}}{80}\log L_{N,\kappa}.
	\label{eq:app-dense-final}
\end{equation}

\subsection{A Single Block with Small Total Pairing Weight}

We now consider the remaining case $L_{N,\kappa}=N$.
By Eq.~\eqref{eq:app-global-L}, this implies $N\leq\lfloor(2r_{N,\kappa}^{2})^{-1}\rfloor$, and hence $Nr_{N,\kappa}^{2}\leq1/2$.
Taking $G=\{0,\ldots,N-1\}$, the corresponding block state is the full embedded pure state, $\rho_{G} = \ket{\widetilde{\Omega}_{N,\kappa}} \bra{\widetilde{\Omega}_{N,\kappa}}$.
For this block, Eq.~\eqref{eq:block-reference-and-probability} gives the success probability of applying the projection $R$ to the reference state $\ket{\Phi_{r_{N,\kappa},G}}$:
\begin{equation}
	p_{N}(r_{N,\kappa}) = \frac{Nr_{N,\kappa}^{2}}{(1+r_{N,\kappa}^{2})^{N}}.
	\label{eq:app-sparse-probability}
\end{equation}
Since $(1+r_{N,\kappa}^{2})^{N}\leq\ee^{Nr_{N,\kappa}^{2}}$ and $Nr_{N,\kappa}^{2}\leq1/2$, we obtain
\begin{align}
	p_{N}(r_{N,\kappa})
	&\geq Nr_{N,\kappa}^{2}\ee^{-Nr_{N,\kappa}^{2}} \notag \\
	&\geq\ee^{-1/2}Nr_{N,\kappa}^{2} \notag \\
	&>0.60Nr_{N,\kappa}^{2}.
	\label{eq:app-sparse-p}
\end{align}
Thus, the success probability of this projection is bounded below by a fixed positive multiple of $Nr_{N,\kappa}^{2}$.

We next verify the condition in Eq.~\eqref{eq:block-good-condition} of Proposition~\ref{prop:block-extraction}.
For the full block, the total excitation weight is
\begin{equation}
	\epsilon_{G} = \sum_{b=0}^{N-1} \bra{\widetilde{\Omega}_{N,\kappa}} Q_{b}(r_{N,\kappa}) \ket{\widetilde{\Omega}_{N,\kappa}}.
	\label{eq:app-sparse-defect}
\end{equation}
Applying Eq.~\eqref{eq:app-reference-defect} at $g=g_{N,\kappa}$ gives $\epsilon_{G}\leq11Nr_{N,\kappa}^{4}$.
To compare the upper bound on $\epsilon_{G}$ with $p_{N}(r_{N,\kappa})^{2}$, we rewrite $11Nr_{N,\kappa}^{4}$ as $11(Nr_{N,\kappa}^{2})^{2}/N$.
The assumption $N\geq4096$ ensures that $11/N\leq11/4096<0.60^{2}/100$.
Together with the probability lower bound in Eq.~\eqref{eq:app-sparse-p}, this gives
\begin{align}
	\epsilon_{G} \leq\frac{11}{N}(Nr_{N,\kappa}^{2})^{2} \leq\frac{p_{N}(r_{N,\kappa})^{2}}{100}.
	\label{eq:app-sparse-good}
\end{align}
Therefore, the full block satisfies the excitation condition in Eq.~\eqref{eq:block-good-condition}.
It also satisfies the size condition $L_{N,\kappa} = N \geq 4096$ and the coupling condition $r_{N,\kappa} \leq 1/128$ required by Proposition~\ref{prop:block-extraction}.

We can therefore apply Proposition~\ref{prop:block-extraction} directly to $\rho_{G}$.
Because $G$ contains all reference channels, its squashed entanglement equals the entanglement entropy of the original ground state by Eq.~\eqref{eq:app-global-pure-sq}.
Using the lower bound on $p_{N}(r_{N,\kappa})$ in Eq.~\eqref{eq:app-sparse-p}, we obtain
\begin{align}
	S_{A}(\ket{\Omega_{N,\kappa}})
	&\geq\frac{9}{50}p_{N}(r_{N,\kappa})\log N \notag \\
	&>0.108Nr_{N,\kappa}^{2}\log N \notag \\
	&\geq\frac{Nr_{N,\kappa}^{2}}{80}
	\log L_{N,\kappa}.
	\label{eq:app-sparse-final}
\end{align}
Here the last inequality uses $L_{N,\kappa}=N$ and $0.108>1/80$.

The lower bounds in Eqs.~\eqref{eq:app-dense-final} and \eqref{eq:app-sparse-final} cover the two possible cases $L_{N,\kappa}<N$ and $L_{N,\kappa}=N$.
Thus, every normalized pure ground state of $H_{N,\kappa}$ satisfies the entropy lower bound in Theorem~\ref{thm:finite-size-main}.

\section{Detailed Weak-Coupling Calculations}
\label{app:weak-coupling}

This appendix derives the Gaussian entropy formulas and the fixed-size spin perturbation expansion used in Sec.~\ref{sec:weak-coupling}.

\subsection{Two-Mode Bogoliubov Diagonalization}

We consider one of the identical two-mode blocks in Eq.~\eqref{eq:HP}, suppressing the channel index:
\begin{equation}
	h_{\mathrm{HP}}(g)=d^{\dagger} d+b^{\dagger} b-g(d^{\dagger} b^{\dagger}+db).
	\label{eq:app-HP-block}
\end{equation}
For $0\leq g<1$, choose $\theta\geq0$ such that $\tanh(2\theta)=g$ and set $\lambda(g)=\tanh\theta$.
Then $\lambda(g)=g/(1+\sqrt{1-g^{2}})$, as in Eq.~\eqref{eq:HP-parameters}.
Define canonical annihilation operators
\begin{align}
	\alpha&:=\cosh\theta\,d-\sinh\theta\,b^{\dagger},
	\label{eq:app-Bogoliubov-alpha}\\
	\beta&:=\cosh\theta\,b-\sinh\theta\,d^{\dagger}.
	\label{eq:app-Bogoliubov-beta}
\end{align}
Their inverse transformation is $d=\cosh\theta\,\alpha+\sinh\theta\,\beta^{\dagger}$ and $b=\cosh\theta\,\beta+\sinh\theta\,\alpha^{\dagger}$.
Substituting the inverse transformations into Eq.~\eqref{eq:app-HP-block}, we obtain
\begin{equation}
	h_{\mathrm{HP}}(g)=\varepsilon(g)(\alpha^{\dagger}\alpha+\beta^{\dagger}\beta)+\varepsilon(g)-1,
	\label{eq:app-HP-diagonal}
\end{equation}
where $\varepsilon(g)=\sqrt{1-g^{2}}$.
The simultaneous vacuum of $\alpha$ and $\beta$ is
\begin{equation}
	\ket{\omega_{\mathrm{HP}}} =\sqrt{1-\lambda(g)^{2}}\sum_{m=0}^{\infty}\lambda(g)^{m}\ket{m}_{d}\ket{m}_{b}.
	\label{eq:app-squeezed-vacuum}
\end{equation}
Indeed, $(d-\lambda(g)b^{\dagger})\ket{\omega_{\mathrm{HP}}}=0$ and $(b-\lambda(g)d^{\dagger})\ket{\omega_{\mathrm{HP}}}=0$, and the geometric series fixes the normalization.
The tensor product of these states over $N$ channels proves Eq.~\eqref{eq:HP-ground-state}.

\subsection{Gaussian Entropy and Its Expansion}

Tracing out the $b$ mode from the two-mode squeezed state in Eq.~\eqref{eq:app-squeezed-vacuum} gives the reduced density operator of the $d$ mode,
\begin{align}
	\rho_{d} &:=\Tr_{b}\br{\ket{\omega_{\mathrm{HP}}}\bra{\omega_{\mathrm{HP}}}} \notag \\
	&=(1-\lambda(g)^{2}) \sum_{m=0}^{\infty}\lambda(g)^{2m}\ket{m}_{d}\bra{m}_{d}.
	\label{eq:app-HP-reduced-state}
\end{align}
Using the identity $\sum_{m=0}^{\infty}mx^{m}=x/(1-x)^{2}$ for $0\leq x<1$, the mean occupation number is
\begin{equation}
	\nu(g)=\frac{\lambda(g)^{2}}{1-\lambda(g)^{2}} = \frac{1}{2}\br{\frac{1}{\sqrt{1-g^{2}}}-1}.
	\label{eq:app-HP-occupation}
\end{equation}
Since $\rho_{d}$ is diagonal in the occupation-number basis, its von Neumann entropy is obtained by summing $-(1-\lambda(g)^{2})\lambda(g)^{2m}\log\br{(1-\lambda(g)^{2})\lambda(g)^{2m}}$ over $m\geq0$.
Using the normalization of the geometric distribution and its mean occupation number $\nu(g)$, we obtain
\begin{equation}
	S(\rho_{d})=-\log(1-\lambda(g)^{2})-\nu(g)\log\lambda(g)^{2}.
	\label{eq:app-geometric-entropy-intermediate}
\end{equation}
The relations $\lambda(g)^{2}=\nu(g)/(1+\nu(g))$ and $1-\lambda(g)^{2}=1/(1+\nu(g))$ then give
\begin{equation}
	S(\rho_{d})=(\nu(g)+1)\log(\nu(g)+1)-\nu(g)\log\nu(g).
	\label{eq:app-geometric-entropy}
\end{equation}
Since the Fourier rotation acts only on subsystem $A$, it preserves the $A|B$ entanglement.
Summing the entropies of the $N$ independent two-mode squeezed states yields Eq.~\eqref{eq:HP-entropy-exact}.

For small $g$,
\begin{equation}
	\nu(g)=\frac{1}{4}g^{2}+\frac{3}{16}g^{4}+\frac{5}{32}g^{6}+\cO(g^{8}).
	\label{eq:app-nu-expansion}
\end{equation}
Expanding the entropy function at small occupation gives $(1+\nu)\log(1+\nu)-\nu\log\nu=\nu(1-\log\nu)+\nu^{2}/2-\nu^{3}/6+\cO(\nu^{4})$.
Substituting Eq.~\eqref{eq:app-nu-expansion} then gives
\begin{align}
	S_{A}^{\mathrm{HP}}(g) &= N\Biggl\{\frac{g^{2}}{4} \left[1-\log\left(\frac{g^{2}}{4}\right)\right] \notag\\
	&\qquad +g^{4}\left[\frac{1}{32} - \frac{3}{16}\log\left(\frac{g^{2}}{4}\right)\right] \notag\\
	&\qquad +g^{6}\left[-\frac{5}{192}
	-\frac{5}{32}\log\left(\frac{g^{2}}{4}\right)\right]\Biggr\} \notag\\
	&\qquad +\cO\bigl(Ng^{8}|\log g|\bigr).
	\label{eq:app-HP-entropy-series}
\end{align}
In particular, the leading term and remainder are
\begin{equation}
	S_{A}^{\mathrm{HP}}(g) = \frac{Ng^{2}}{4} \left[1-\log\left(\frac{g^{2}}{4}\right)\right]  + \cO\bigl(Ng^{4}|\log g|\bigr).
	\label{eq:app-HP-entropy-leading}
\end{equation}
For fixed $0 < \gamma \leq 1/64$ and $1/2 < \kappa \leq 1$, inserting $g_{N,\kappa}=\gamma N^{\frac{1}{2}-\kappa}$ gives Eq.~\eqref{eq:HP-entropy-N}.
At $\kappa = 1/2$, one must instead keep $g_{N,\kappa}=\gamma$ fixed and use the exact Gaussian entropy per channel, giving $S_{A}^{\mathrm{HP}}=Ns_{\mathrm{HP}}(\gamma)$.

\subsection{Ground Energy in Fixed-Size Spin Perturbation Theory}

Fix $N$ and consider the perturbative expansion of $H_{N}(g)$ around $g=0$.
At $g=0$, $\ket{\mathbf{0}}$ is the unique ground state and is separated from the first excited cluster by energy $1$.
Finite-dimensional perturbation theory therefore applies for sufficiently small $|g|$~\cite{Kato1995}.

The first-order energy correction vanishes because the pairing operators $P_{N}^{\dagger}$ and $P_{N}$ change the total excitation number by $+2$ and $-2$, respectively.
In second-order perturbation theory, the coefficient of $g^{2}$ in the ground-state energy expansion is obtained by inserting the resolvent of the unperturbed Hamiltonian between the pairing operators.
Since $P_{N}\ket{\mathbf{0}}=0$, only pair creation followed by pair annihilation contributes, giving
\begin{equation}
	E_{N}^{(2)} = \bra{\mathbf{0}}P_{N} \frac{1}{0-\mathcal{N}} P_{N}^{\dagger} \ket{\mathbf{0}}.
	\label{eq:app-spin-energy-resolvent}
\end{equation}
Here the inverse acts on the subspace orthogonal to the vacuum.

The operator $P_{N}^{\dagger}$ connects the vacuum to the two-excitation states $\sigma_{i_{a}}^{+}\sigma_{j_{b}}^{+}\ket{\mathbf{0}}$, each with unperturbed energy $2$.
The corresponding transition amplitudes are $(F_{N})_{ab}$, whose total squared magnitude is
\begin{equation}
	\bra{\mathbf{0}}P_{N}P_{N}^{\dagger} \ket{\mathbf{0}} = \sum_{a,b}|(F_{N})_{ab}|^{2}=N.
	\label{eq:app-vacuum-pair-norm}
\end{equation}
Since all contributing intermediate states have energy $2$, the energy denominator in Eq.~\eqref{eq:app-spin-energy-resolvent} is $0-2=-2$.
Consequently,
\begin{equation}
	E_{N}^{(2)} = \sum_{a,b}\frac{|(F_{N})_{ab}|^{2}}{0-2} = -\frac{N}{2}.
	\label{eq:app-spin-energy-second-order}
\end{equation}

Conjugation by $\ee^{\ii\pi\mathcal{N}_{A}}$ leaves $\mathcal{N}$ invariant and changes the signs of $P_{N}^{\dagger}$ and $P_{N}$, so $H_{N}(g)$ and $H_{N}(-g)$ are unitarily equivalent.
Consequently, $E_{N,0}(g) = E_{N,0}(-g)$.
Since $E_{N,0}(g)$ is analytic near $g=0$ for fixed $N$, its Taylor expansion contains only even powers of $g$.
Thus,
\begin{equation}
	E_{N,0}(g)=-\frac{N}{2}g^{2}+\cO_{N}(g^{4}).
	\label{eq:app-spin-energy-fixedN}
\end{equation}

\subsection{The One-Spin Excited Cluster}

At $g=0$, the one-spin excited states form a $2N$-dimensional eigenspace of $H_{N}(0)=\mathcal{N}$ with energy $1$.
Let $Q_{A}^{(1)}$ denote the projector onto the subspace with exactly one excited spin in $A$ and no excitations in $B$, and let $Q_{B}^{(1)}$ denote the analogous projector with exactly one excited spin in $B$ and no excitations in $A$.
The projector $\mathscr{P}_{1}$ introduced in Sec.~\ref{subsec:spin-gap-perturbation} satisfies $\mathscr{P}_{1}=Q_{A}^{(1)}+Q_{B}^{(1)}$.
Since the pairing operators change the total excitation number by two, $\mathscr{P}_{1}(P_{N}^{\dagger}+P_{N})\mathscr{P}_{1}=0$, so the first-order perturbative correction vanishes throughout this manifold.

To determine the second-order correction, we first evaluate the matrix elements of $P_{N}P_{N}^{\dagger}$ within the one-spin manifold.
Consider the state $\ket{c}_{A}:=\sigma_{i_{c}}^{+}\ket{\mathbf{0}}$, where $c=0,\ldots,N-1$.
Applying $P_{N}^{\dagger}$ to this state gives
\begin{equation}
	P_{N}^{\dagger}\ket{c}_{A}
	=\sum_{\substack{a=0\\a\neq c}}^{N-1}
	\sum_{b=0}^{N-1}(F_{N})_{ab}
	\sigma_{i_{a}}^{+}\sigma_{j_{b}}^{+}\ket{c}_{A}.
	\label{eq:app-blocked-pair-vector}
\end{equation}
The terms with $a=c$ vanish because $(\sigma_{i_{c}}^{+})^{2}=0$, reflecting the hard-core constraint on the already excited spin.
The resulting three-excitation computational-basis states are mutually orthogonal.
Using the normalization of each row of $F_{N}$, we therefore obtain
\begin{equation}
	\bra{c}_{A} P_{N}P_{N}^{\dagger}\ket{c}_{A} = \sum_{a\neq c,b}|(F_{N})_{ab}|^{2} = N-1.
	\label{eq:app-blocked-pair-norm}
\end{equation}
Thus, the occupied spin removes one unit of pair-creation weight from the vacuum value $N$ in Eq.~\eqref{eq:app-vacuum-pair-norm}.

We next evaluate the off-diagonal matrix elements.
For $c\neq c'$, a common intermediate state contributing to the overlap of $P_{N}^{\dagger}\ket{c}_{A}$ and $P_{N}^{\dagger}\ket{c'}_{A}$ must be obtained by creating excitations at sites $i_{c'}$ and $j_{b}$ from $\ket{c}_{A}$, and at sites $i_{c}$ and $j_{b}$ from $\ket{c'}_{A}$.
The corresponding amplitudes are $(F_{N})_{c'b}$ and $(F_{N})_{cb}$, respectively.
Thus,
\begin{equation}
	\bra{c'}_{A} P_{N}P_{N}^{\dagger}\ket{c}_{A}
	=\sum_{b}\overline{(F_{N})_{cb}}(F_{N})_{c'b}
	=0,
	\label{eq:app-oneA-offdiagonal}
\end{equation}
where the last equality follows from the orthogonality of distinct rows of $F_{N}$.
Together with Eq.~\eqref{eq:app-blocked-pair-norm}, this shows that $P_{N}P_{N}^{\dagger}$ is proportional to the identity on the one-$A$-spin subspace:
\begin{equation}
	Q_{A}^{(1)}P_{N}P_{N}^{\dagger}Q_{A}^{(1)} = (N-1)Q_{A}^{(1)}.
	\label{eq:app-oneA-identity}
\end{equation}

The same calculation applies to the one-$B$-spin subspace.
For a state with its only excitation at site $j_{c}$, the terms with $b=c$ in $P_{N}^{\dagger}$ vanish because $(\sigma_{j_{c}}^{+})^{2}=0$.
Column normalization gives the diagonal matrix element $N-1$, while column orthogonality makes the off-diagonal matrix elements vanish.
Consequently,
\begin{equation}
	Q_{B}^{(1)}P_{N}P_{N}^{\dagger}Q_{B}^{(1)} = (N-1)Q_{B}^{(1)}.
	\label{eq:app-oneB-identity}
\end{equation}
Moreover, the one-$A$-spin and one-$B$-spin subspaces have eigenvalues $+1$ and $-1$, respectively, under the conserved number-difference operator $\mathcal{D}$.
Since $P_{N}$ and $P_{N}^{\dagger}$ commute with $\mathcal{D}$, the matrix elements of $P_{N}P_{N}^{\dagger}$ between these two subspaces vanish.
Combining Eqs.~\eqref{eq:app-oneA-identity} and \eqref{eq:app-oneB-identity}, we conclude that
\begin{equation}
	\mathscr{P}_{1}P_{N}P_{N}^{\dagger}\mathscr{P}_{1} = (N-1)\mathscr{P}_{1}.
	\label{eq:app-entire-one-spin-identity}
\end{equation}

We now apply second-order perturbation theory to the degenerate one-spin manifold.
Since $P_{N}$ annihilates every one-spin state, only pair creation followed by pair annihilation contributes.
The intermediate states have three excitations and therefore unperturbed energy $3$, whereas the one-spin manifold has energy $1$.
The second-order effective operator is consequently
\begin{align}
	\mathscr{P}_{1}P_{N}\frac{1}{1-\mathcal{N}}P_{N}^{\dagger}\mathscr{P}_{1} = -\frac{1}{2}\mathscr{P}_{1}P_{N}P_{N}^{\dagger}\mathscr{P}_{1}   = -\frac{N-1}{2}\mathscr{P}_{1}.
	\label{eq:app-second-order-odd-operator}
\end{align}
Here the inverse of $1-\mathcal{N}$ acts on the three-excitation subspace reached by $P_{N}^{\dagger}\mathscr{P}_{1}$, where it equals $-1/2$.
The last equality follows from Eq.~\eqref{eq:app-entire-one-spin-identity}.

Since the second-order effective operator is proportional to $\mathscr{P}_{1}$, all $2N$ eigenvalues emerging from the one-spin manifold receive the same second-order correction.
Furthermore, an odd number of pair-creation and pair-annihilation operations cannot return a state to its original total-excitation sector, so all odd-order corrections vanish.
For fixed $N$ and sufficiently small $|g|$, this cluster remains the first excited cluster, yielding
\begin{equation}
	E_{N,1}(g)=1-\frac{N-1}{2}g^{2}+\cO_{N}(g^{4}).
	\label{eq:app-one-spin-cluster}
\end{equation}

Combining Eqs.~\eqref{eq:app-one-spin-cluster} and \eqref{eq:app-spin-energy-fixedN}, we obtain
\begin{equation}
	\Delta_{N}(g) = 1 + \frac{1}{2}g^{2} + \cO_{N}(g^{4}).
	\label{eq:app-spin-gap-final}
\end{equation}
This proves Eq.~\eqref{eq:spin-gap-perturbative} in the main text.

\end{document}